\documentclass[a4paper,12pt]{article}
\usepackage{latexsym}
\usepackage{amsfonts}
\usepackage{amsmath}
\usepackage{amsthm}
\usepackage{amssymb}
\usepackage{hyperref}
\usepackage{graphicx}

\def\d{\partial}

\newcommand{\dl}[1]{\displaystyle\frac{{\d}}{\d #1}}

\def\half{{\frac{1}{2}}}

\numberwithin{equation}{section}
\def\be{\begin{eqnarray}}
\def\ee{\end{eqnarray}}
\def\beann{\begin{eqnarray*}}
\def\eeann{\end{eqnarray*}}
\def\beq{\begin{equation}}
\def\eeq{\end{equation}}
\def\ba{\begin{array}}
\def\ea{\end{array}}
\def\ben{\begin{enumerate}}
\def\een{\end{enumerate}}
\def\bea{\begin{eqnarray}}
\def\eea{\end{eqnarray}}
\def\beann{\begin{eqnarray*}}
\def\eeann{\end{eqnarray*}}
\def\beq{\begin{equation}}
\def\eeq{\end{equation}}
\def\ba{\begin{array}}
\def\ea{\end{array}}
\def\ben{\begin{enumerate}}
\def\een{\end{enumerate}}

\def\5{\bar }
\def\6{\partial }
\def\7{\hat }
\def\4{\tilde }

\def\cA{{\cal A}}
\def\cB{{\cal B}}
\def\cC{{\cal C}}
\def\cD{{\cal D}}
\def\cE{{\cal E}}
\def\cF{{\cal F}}
\def\cG{{\cal G}}

\def\cI{{\cal I}}
\def\cJ{{\cal J}}
\def\cK{{\cal K}}

\def\cM{{\cal M}}
\def\cN{{\cal N}}

\def\cP{{\cal P}}

\def\s0#1#2{\mbox{\small{$\frac{#1}{#2}$}}}

\newtheorem{theorem}{Theorem}
\newtheorem{corollary}{Corollary}
\newtheorem{lemma}{Lemma}
\def\qed{\hbox{${\vcenter{\vbox{
\hrule height 0.4pt\hbox{\vrule width 0.4pt height 6pt
\kern5pt\vrule width 0.4pt}\hrule height 0.4pt}}}$}}

\begin{document}

\begin{titlepage}

\begin{centering}

\vspace{0.5cm}

\huge{Results on the Wess-Zumino consistency condition for
arbitrary Lie algebras}  \\

\vspace{.5cm}

\large{A.~Barkallil, G.~Barnich$^{*}$ and C.~Schomblond
}

\vspace{.5cm}

Physique Th\'eorique et Math\'ematique,\\ Universit\'e Libre de
Bruxelles,\\
Campus Plaine C.P. 231, B--1050 Bruxelles, Belgium

\vspace{.5cm}

\end{centering}

\begin{abstract}
The so-called covariant Poincar\'e lemma on the induced cohomology of the
spacetime exterior derivative in the cohomology of the gauge part of
the BRST differential is extended to cover the case of arbitrary, non reductive
Lie algebras. As a consequence, the general solution of the
Wess-Zumino consistency condition with a non trivial descent can, for
arbitrary (super) Lie algebras, be computed in the small algebra of the 1 form
potentials, the ghosts and their exterior derivatives.
For particular Lie algebras that are the
semidirect sum of a semisimple Lie subalgebra with an ideal, a
theorem by Hochschild and Serre is used to characterize more precisely
the cohomology of the gauge part of the BRST differential in the small
algebra. In the case of an abelian ideal, this leads to a complete
solution of the
Wess-Zumino consistency condition in this space. As
an application, the consistent deformations of 2+1 dimensional
Chern-Simons theory based on iso(2,1) are rediscussed.
\end{abstract}

\vspace{.5cm}

\footnotesize{$^*$ Research Associate of the National Fund for
  Scientific Research (Belgium).}
\vfill

\end{titlepage}

\section{Introduction}

The algebraic problem that is central for the renormalization of Yang-Mills
theory is the computation of $H^{0,n}(s|d)$ and $H^{1,n}(s|d)$, the
cohomology of the BRST differential modulo the exterior spacetime
differential $d$ in ghost number $0$ and $1$ in the space of fields,
external sources (called antifields below) and their
derivatives \cite{Becchi:1975nq} (see also e.g.
\cite{Dixon:1979bs} and \cite{Piguet:1995er,Bonneau:1990xu} for reviews).

So far, local BRST cohomology groups for Yang-Mills and Chern-Simons theories
have been investigated exclusively in the context of reductive Lie
algebras, i.e., Lie algebras that are the direct sum of a semisimple and
abelian factors. In this case, the invariant metric used in the construction of
the actions is necessarily the Killing metric for the semisimple factor (up to
an overall constant), complemented by an arbitrary metric  for the abelian
factors. Recently, there has been a lot of interest in non reductive Lie
algebras that nevertheless possess an invariant metric
\cite{Figueroa-O'Farrill:1994yf}, as it is then still possible to construct
Wess-Zumino-Witten models, Chern-Simons and Yang-Mills theories (see
\cite{Tseytlin:1995yw} and references therein). In particular, the associated
Yang-Mills theories have remarkable renormalization properties. This motivates
the study of the local BRST cohomology groups for such theories.

An important intermediate step in this study is the computation of the local
BRST cohomology $H(\gamma|d)$ of the gauge part $\gamma$ of the BRST
differential in the algebra ${\cal A}$ generated by the spacetime forms, the
gauge potentials, ghosts and a finite number of their derivatives. This
computation in turn relies crucially on the so called covariant
Poincar\'e lemma
\cite{Brandt:1990gy,Brandt:1990et,Dubois-Violette:1992ye}.
In the reductive case, the covariant Poincar\'e lemma states that
the cohomology $H(d,H(\gamma,{\cal A}))$ is generated by the invariant
polynomials in the curvature $2$-forms $F^a$ and invariant polynomials in
the ghosts $C^a$. The proof of this lemma uses the fact that
for reductive Lie algebras, the (Chevalley-Eilenberg) Lie algebra cohomology
\cite{Chevalley} with coefficients in a finite dimensional module $V$ is
isomorphic to the tensor product of the invariant subspace of the module
(Whitehead's theorem) and the Lie algebra cohomology with coefficients in
$\mathbb R$, which is itself generated by the primitive elements (see e.g.
\cite{Postnikov,Greub} and also \cite{Brandt:1990gv}).
The consequence for the computation of
$H(\gamma|d)$ is that all the solutions of the Wess-Zumino consistency
condition\cite{Wess:1971yu} with a non trivial descent can be computed
in the small
algebra ${\cal B}$ generated by the $1$-form potentials, the ghosts and their
exterior derivatives, thus providing an a posteriori justification of the
assumptions of
\cite{Dubois-Violette:1985hc,Dubois-Violette:1985jb,Talon:1985dz%
,Dubois-Violette:1986cj} (see also
\cite{Stora:1983ct,Zumino:1984rz,Zumino:1983ew,Manes:1985df} for related
considerations).

The central result of the present paper is the generalization of the
covariant Poincar\'e lemma to arbitrary Lie algebras
for which the Lie algebra cohomology is not necessarily explicitly
known. In order to do so, we will use a
standard decomposition according to the
homogeneity in the fields.
From the proof, it will also be obvious that the result extends to the case of
super/graded Lie algebras.

For particular Lie algebras $\cG$ that admit an ideal $\cJ$ such that
the quotient $\cG/\cJ$
is semisimple, we use a theorem by Hochschild and Serre
\cite{HochschildSerre} that states that the Lie algebra cohomology of
$\cG$  with coefficients in $V$ reduces to the tensor product of the
Lie algebra cohomology of the semisimple factor $\cG/\cJ$ with
coefficients
$\mathbb R$ and the invariant cohomology of the ideal $\cJ$ with
coefficients in $V$. As this result is not as widely known as the
standard results on reductive Lie algebra  cohomology, we will
rederive it using "ghost" language, i.e., by writing the cochains with
coefficients in $V$ as polynomials in the Grassmann odd generators
$C^a$ with coefficients in $V$ and by writing the Chevalley-Eilenberg
differential as a first order differential operator acting in this
space. As a direct application, we explicitly compute $H(\gamma,\cB)$
for the three dimensional Euclidian and Poincar\'e algebras $iso(3)$
and $iso(2,1)$.

In the case where the ideal $\cJ$ is abelian, this leads to a complete
characterization of $H(\gamma|d,{\cal B})$, allowing us in particular
to give exhaustive results for $iso(3)$ and $iso(2,1)$. The
covariant  Poincar\'e lemma, allows to extend these results to
$H(\gamma|d,{\cal A})$.

Finally, we explicitly rediscuss the local BRST cohomology, and more
particularly the consistent deformations, of $iso(2,1)$
Chern-Simons theory, whose physical relevance is due to its relation
with $2+1$ dimensional gravity \cite{Achucarro:1986vz,Witten:1988hc}.

\section{Generalities and conventions}

We take spacetime to be $n$-dimensional Minkowski space with $n\geq 3$ and
${\cal A}$ to be either the algebra of form valued polynomials or
the algebra of form valued formal power series in the
potentials $A^a_\mu$, the ghosts $C^a$ (collectively denoted by $\phi^i$)
and their derivatives. The algebra ${\cal A}$ can be decomposed into subspaces
of definite ghost number $g$, by assigning ghost number $1$ to the ghosts and
their derivatives and ghost number zero to $x^\mu$, $dx^\mu$, the gauge
potentials and their derivatives. Let ${\cal B}$ be the either the
algebra of polynomials or of formal power series
generated by $A^a,C^a,dA^a,dC^a$, with $d=dx^\mu\partial_\mu$ and
$\partial_\mu$
denoting the total derivative. Let $f_{ab}^c$ be the structure constants of a
Lie algebra ${\cal G}$. The action of the gauge part $\gamma$ of the BRST
differential is defined by \bea
\gamma A_\mu^a=\partial_\mu C^a+f^a_{bc} A^b_\mu C^c,
\gamma C^a=-\frac{1}{2}f^a_{bc}C^bC^c,\\
\gamma x^\mu=\gamma dx^\mu=0,[\gamma,\partial_\mu]=
\{\gamma,d\}=0. \label{2.2}
\eea
Let $F^a_{\mu\nu}=\partial_\mu A^a_\nu-\partial_\nu
A^a_\mu+f^a_{bc} A^b_\mu A^c_\nu$ so that $\gamma
F^a_{\mu\nu}=f^a_{bc}F^b_{\mu\nu}C^c$.
In ${\cal B}$, the action of $\gamma$ reads
\bea
\gamma A^a=-DC^a,\\
\gamma C^a =-\frac{1}{2}[C,C]^a.
\eea
The field strength $2$-form $F^a=dA^a+\frac{1}{2}[A,A]^a$ satisfies
\bea
\gamma F^a=[F,C]^a,\ D F^a=0,
\eea
with $D=d +[A,\cdot]$.

In the following, the algebra $\cE$ stands for either $\cA$ or $\cB$.
Under the above assumptions, the cohomology of $d$ is known to be trivial in
$\cE$ \cite{Vinogradov:1977,Takens:1979aa,Tulczyjew:1980aa,%
Anderson:1980aa,DeWilde:1981aa,Tsujishita:1982aa,Olver:1993,%
Brandt:1990gy,Wald:1990aa,Dubois-Violette:1991is,Dickey:1992aa}.
More precisely, in form degree $0$, the cohomology of $d$ is exhausted by the
constants, and in particular it is trivial in strictly positive ghost numbers.
It is also trivial in form degrees $0<p<n$.

A standard technique for computing $H(\gamma|d,\cE)$ is to use so
called descent equations. As a consequence of the
acyclicity of the exterior differential $d$,
the cocyle condition $\gamma \omega^p+d\omega^{p-1}=0$
implies that $\gamma \omega^{p-1}
+d\omega^{p-2}=0$. Iterating the descent, there necessarily exists an
equation which reads $\gamma\omega^{p-l}=0$ because the form degree
cannot be lower than zero. One then tries to compute $H(\gamma|d,\cE)$
by starting from the last equation. The systematics of this strategy
can be captured by the exact couple
\bea
\cC=<H(\gamma|d,\cE ),H(\gamma,\cE ),{\cal D},l^{\#},i^{\#}>,
\eea
\begin{eqnarray}
\begin{array}{ccc} H(\gamma|d,\cE )\stackrel{{\cal D}}{\longrightarrow}
H(\gamma|d,\cE )\\ i^{\#}\nwarrow\ \swarrow l^{\#}\\ H(\gamma,\cE )
\end{array}\label{ec}
\end{eqnarray}
and the associated spectral sequence \cite{Dubois-Violette:1985jb}
(see also \cite{Henneaux:1998rp,Barnich:2000zw} for reviews).
The various maps are defined as follows:  $i^{\#}$ is the map which
consists in regarding an element of $H(\gamma,\cE )$ as an element of
$H(\gamma|d,\cE )$,
$i^{\#}:H(\gamma,\cE )\longrightarrow
H(\gamma|d,\cE )$, with $i^{\#}[\omega]=[\omega]$. It is well defined
because every $\gamma$ cocycle is a
$\gamma$ cocycle
modulo $d$ and every $\gamma$ coboundary is a $\gamma$
coboundary modulo $d$.
The descent homomorphism
${\cal D}:
H^{k,l}(\gamma|d,\cE )\longrightarrow
H^{k+1,l-1}(\gamma|d,\cE )$ with
${\cal D}[\omega]=[\omega^\prime]$, if $\gamma \omega + d\omega^\prime=0$ is
well defined because of the triviality of the cohomology of $d$ in
form degree $p\leq n-1$ (and ghost number $\geq 1$).
Finally, the map $l^{\#}:H^{k+1,l-1}(\gamma|d,\cE )\longrightarrow
H^{k+1,l}(\gamma,\cE )$ is defined by $l^{\#}[\omega]=[d \omega]$. It is well
defined because the relation $\{\gamma,d\}=0$ implies that
it maps cocycles to cocycles and coboundaries to coboundaries.
The differential associated to the exact triangle is
$d^{\#}=l^{\#}\circ i^{\#}$.

The exactness of the couple \eqref{ec} implies that
\bea
H(\gamma,\cE)\simeq
H(d^{\#},H(\gamma,\cE))\oplus d^{\#}\cN_\cE\oplus \cN_\cE,\label{gammaC}\\
H(\gamma|d,\cE))= i^{\#} H(\gamma,\cE)\oplus \cD^{-1} \cD H
(\gamma|d,\cE)\label{gammadC},
\eea
where $ \cN_\cE$ is the subspace of  $H(\gamma,\cE))$ which cannot be lifted,
i.e., $[c]\in  \cN_\cE$ if $\gamma c=0$ with $dc+\gamma\omega=0\Longrightarrow
c=\gamma\omega'$.

In the following, $\omega,a,A\in \cA$ and $\varpi,b,B\in \cB$.

\section{The first lift in the small algebra\label{s3}}

In $\cB$, the change of generators from $A^a,dA^a,C^a,dC^a$ to
$A^a,F^a,C^a,-DC^a$ allows to isolate the contractible pairs $A^a,-DC^a$
and the cohomology of $\gamma$ can be computed in the polynomial algebra
generated by $F^a,C^a$. In other words, $\gamma b=0\Longleftrightarrow b=
P(F,C)+\gamma b'$ with $\gamma P(F,C)=0$, while $P(F,C)=\gamma
b'\Longrightarrow P(F,C)=\gamma P'(F,C)$.
Let us first prove the following lemma:
\begin{lemma}  \label{lem1}
Every element of $H(\gamma,\cB)$ can be lifted at
least once and furthermore, no element of $H(\gamma,\cB)$ is an
obstruction to a
lift of an element of $H(\gamma,\cB)$:
\bea
H^p(d^{\#},H(\gamma,\cB))\simeq
H^p(\gamma,\cB),\ {\rm for}\ 0\leq p\leq n,
\eea
i.e.,
\bea \gamma b^p=0\Longrightarrow
d b^p+\gamma \varpi =0, \label{3.1}
\eea
together with
\bea
\left\{\begin{array}{c}
b^p=d\bar b+\gamma b^\prime,\\
\gamma \bar b=0
\end{array}\right.
\Longrightarrow b^p=\gamma \varpi'. \label{3.2}
\eea
\end{lemma}
\proof{In terms of the new generators, we have
\bea
\gamma=-DC^a\frac{\partial}{\partial A^a}+[F,C]^a\frac{\partial}{\partial
F^a}-\half[C,C]^a\frac{\partial}{\partial C^a},\\
d=(F-\frac{1}{2}[A,A])^a \frac{\partial}{\partial A^a}-[A,F]^a\dl{F^a}
\nonumber\\+(DC-[A,C])^a
\frac{\partial}{\partial
C^a}+([F,C]-[A,DC])^a\dl{DC^a}.\eea
Let us introduce the operator \cite{Brandt:1990gy,Sorella:1993dr}
\bea
\lambda=A^a
\frac{\partial}{\partial C^a}-(F^a-\frac{1}{2}[A,A]^a)
\frac{\partial}{\partial
  DC^a}.
\eea
We get
\bea
d=[\lambda,\gamma].\label{3.7}
\eea
It follows that
\bea
db+\gamma\lambda b=0,
\label{3.8}
\eea
if $\gamma b=0$. Furthermore, if $b=d\bar b+\gamma \varpi$ with $\gamma\bar
b=0$, we get $b=\gamma(\varpi-\lambda \bar b)$.}

If $\gamma b=0$, we also get
\bea
d\lambda b+\gamma\frac{1}{2}\lambda^2 b=\tau b,
\label{3.9}
\eea
with
\bea
\tau=\frac{1}{2}[d,\lambda]=
F^a\frac{\partial}{\partial
C^a},\label{3.10}\\
\tau^2=0,\ \{\tau,\gamma\}=0.\label{3.11}
\eea
It follows that the potential obstructions to lifts of elements of
$H(\gamma,\cB)$ are controlled by the differential $\tau$.

\section{Covariant Poincar\'e lemma for generic (super)-Lie algebras}

\subsection{Formulation}

\begin{theorem}[Covariant Poincar\'e lemma]\label{t1}
The following isomorphism holds:
\bea
H^p(d^{\#},H(\gamma,\cA)) &\simeq& H^p(\gamma,\cB),\ 0\leq p<n.\label{covP}
\eea
\end{theorem}
Explicitly, this means that
\bea
\left\{\begin{array}{c} \gamma a^p=0,\\
d a^p+\gamma \omega=0,\ 0\leq p<n
\end{array}\right.\Longleftrightarrow
\left\{\begin{array}{c} a^p=b^p+d a^{\prime p} +\gamma \omega',\\
\gamma b^p=0=\gamma
a^{\prime p},\end{array}\right.\label{4.5}
\eea
together with
\bea
\left\{\begin{array}{c} b^p+d a^{\prime p} +\gamma \omega'=0,\\
\gamma b^p=0=\gamma a^{\prime p},\ 0\leq p<n\end{array}\right.
\Longrightarrow
 b^p=\gamma \varpi. \label{4.6s}
\eea
In fact, we will prove that \eqref{4.6s} also holds in form degree
$p=n$.

\subsection{Associated structure of $\cD H(\gamma|d,\cA)$}\label{s42}

As a direct consequence of the covariant Poincar\'e lemma, the
descent homomorphism in $\cA$ reduces to that in $\cB$.
\begin{corollary}\label{c1}
The isomorphism \eqref{covP} implies
\bea
(\cD H)^p (\gamma|d,\cA)\simeq (\cD H)^p(\gamma|d,\cB)\label{desc}, \
0\leq p<n.
\eea
More precisely, \eqref{covP} for $0\leq p\leq m(<n)$ implies
\eqref{desc} for $0\leq p\leq m(<n)$.
\end{corollary}
This last isomorphism is equivalent to
\bea
\left\{\begin{array}{c} \gamma A^p+ d A^{\prime p-1} =0,\\
d A^p+\gamma \omega=0 ,\ 0\leq p<n\end{array}\right. \Longleftrightarrow
\left\{\begin{array}{c} A^p=
B^p + \gamma \omega' + d\omega'',\\
\gamma B^p + d B^{\prime p-1}=0,\\
dB^{p}+\gamma B''=0, \end{array}\right. \label{4.7s}
\eea
together with
\bea
\left\{\begin{array}{c}
B^p + \gamma \omega' + d\omega''=0,\\
dB^p+\gamma B''=0,\ 0\leq p<n  \end{array}\right.
\Longrightarrow  B^p=\gamma \varpi+ d\varpi'. \label{4.8s}
\eea

\proof{That the condition \eqref{4.7s} is necessary ($\Longleftarrow$)
follows directly from the properties of $\gamma$ cocycles in the small algebra
proved in section \ref{s3} and the fact that $d$ and $\gamma$
anticommute.

The proof that the condition \eqref{4.7s} is also sufficient and the
proof of \eqref{4.8s} proceeds by induction on the form degree.
In form degree $0$, \eqref{4.7s} and \eqref{4.8s}
hold, because \eqref{4.7s} coincides with \eqref{4.5}, while \eqref{4.8s}
coincides with \eqref{4.6s}. Suppose  \eqref{4.7s} and \eqref{4.8s}
hold in form degree $0\leq p\leq m-1$.

For \eqref{4.7s}, it follows by induction that
$A^{\prime m-1}=B^{\prime m-1}+\gamma\omega''+d(\ )$. This implies $\gamma
(A^m-d\omega'')+dB^{\prime m-1}=0$. This gives $dB^{\prime m-1} +\gamma \bar
B^m=0$
and then
$A^m=B^m+\bar a^m+ d\omega''$ with $\gamma \bar a^m=0$.
The assumption on $A^m$ implies that
$d\bar B^m+d\bar a^m+\gamma \omega=0$. From \eqref{4.6s}, we then deduce that
$d\bar B^m+\gamma \bar B''=0$, and also that $d\bar a^m
+\gamma(\omega-\bar B'')=0$.
Using \eqref{4.5}, we get $\bar a^m = b^m+ d a' +\gamma(\ )$. Hence, because
of \eqref{3.1}, the right hand side \eqref{4.7s} holds with $B^m=\bar B^m+b^m$.

For \eqref{4.8s}, we note that the assumptions imply that
$\gamma B^m+ d B^{\prime m-1}=0$.
Furthermore,
$d(B^{\prime m-1}+\gamma\omega'')=0$, so that $B^{\prime m-1}+
\gamma \omega''+d(\ )=0$. By
induction, this means that $B^{\prime m-1}=\gamma\varpi'+d\varpi''$.
This implies that
$\gamma (B^m-d\varpi')=0$ so that $B^m=d\varpi'+ b^m$.
The assumption that $B^m$ can be lifted in the small algebra then
gives $db^m=-\gamma B''$. Using
\eqref{3.2}, this implies that $b^m=\gamma\varpi$ so that
the r.h.s of \eqref{4.8s} holds in form degree $m$.}

\subsection{Associated structure of $H(\gamma,\cA)$ and
$H(\gamma|d,\cA)$\label{scons}}

Taking \eqref{covP} into account, the decomposition \eqref{gammaC} for
$\cE=\cA$ becomes
\bea
H^p(\gamma,\cA)&\simeq& H^p(\gamma,\cB)\oplus d^{\#}\cN^{p-1}_\cA\oplus
\cN^p_\cA,\ 0\leq p< n, \label{gammaA}\\
H^n(\gamma,\cA)&\simeq &\cF^n \oplus d^{\#}\cN^{n-1}_\cA ,
\eea
with $\cF^n \simeq H^n(\gamma,\cA)/   d^{\#}\cN^{n-1}_\cA$.
This is equivalent to
\bea
\gamma a^p=0,\ 0\leq p< n \Longleftrightarrow  \left\{\begin{array}{c} a^p=
b^p +d\tilde a^{p-1} + \tilde a^{\prime p}+\gamma \omega,\\ \gamma b^p=0=
\gamma \tilde
a^{p-1}=\gamma \tilde a^{\prime p},
\end{array}\right.
\eea
together with
\bea
\left\{\begin{array}{c} b^p +d\tilde a^{p-1} +  \tilde a^{\prime p}
+\gamma \omega=0,\ 0\leq p< n,\\ \gamma b^p=0=\gamma \tilde a^{p-1}=\gamma
\tilde a^{\prime p}
\end{array}\right.
\Longrightarrow  \left\{\begin{array}{c}b^p=\gamma \varpi,\\
\tilde a^{p-1} =\gamma
\omega',\\ \tilde a^{\prime p} =  \gamma \omega'',\end{array}\right.,
\eea
in form degrees $0\leq p <n$ and to
\bea
\gamma a^n=0\Longleftrightarrow  \left\{\begin{array}{c} a^n=
\hat a^n +d\tilde a^{n-1} +\gamma \omega,\\ \gamma \hat a^n=0=\gamma
\tilde a^{n-1},\end{array}\right. \label{4.10}
\eea
together with
\bea
\left\{\begin{array}{c} \hat a^n +d\tilde a^{n-1} +\gamma \omega=0,\\
\gamma
\hat a^n=0=\gamma \tilde a^{n-1}
\end{array}\right. \Longrightarrow  \left\{\begin{array}{c}
\hat a^n=\gamma
\omega,\\ \tilde a^{n-1} =\gamma \omega',\end{array}\right.
\eea
in form degree $n$.

Finally, using \eqref{gammaA} and \eqref{desc}, the decomposition
\eqref{gammadC} for $\cE=\cA$ becomes
\bea
H^p(\gamma|d,\cA)&\simeq& i^{\#} H^p(\gamma,\cB)\oplus
\cN^p_\cA \oplus \cD^{-1}
(\cD H)^{p-1} (\gamma|d,\cB), 0\leq p<n,\label{gmoddA}\\
H^n(\gamma|d,\cA)&\simeq & \cF^n\oplus  \cD^{-1} (\cD
H)^{n-1} (\gamma|d,\cB),\label{gmoddAn}
\eea
which is equivalent to
\bea
\gamma A^p+ dA^{\prime p}=0,\ 0\leq p< n\Longleftrightarrow  \left\{\begin{array}{c}
A^p= \tilde a^p +b^p+ B^p +\gamma \omega +d \omega',\\ \gamma \tilde a^p
=0=\gamma b^p=\gamma B^p+ dB^{\prime p-1},\\ d\tilde a^p+\gamma\omega''=0
\Rightarrow
\tilde a^p=\gamma\omega''',\\ B^{\prime p-1}
=\gamma \varpi +d \varpi'\Rightarrow
B^p=\gamma \varpi'' +d\varpi,\end{array}\right. \label{4.13}
\eea
together with
\bea
\left\{\begin{array}{c} \tilde a^p +b^p + B^p
+\gamma \omega +d \omega'=0,\ 0\leq p< n\\ \gamma \tilde a^p =0=\gamma
b^p=
\gamma B^{p}+
dB^{\prime p-1},\\ d\tilde a^p+\gamma\omega''=0\Rightarrow  \tilde
a^p=\gamma\omega''',\\ B^{\prime p-1}=\gamma \varpi +d \varpi'\Rightarrow
B^p=
\gamma
\varpi'' +d\varpi\end{array}\right.
\Longrightarrow  \left\{\begin{array}{c} \tilde a^p = \gamma \omega^4,\\
b^p=dB^{\prime\prime p-1}+\gamma\varpi''',\\ B^p=\gamma \varpi'''+
d\varpi^{4},\end{array}\right. \eea
in form degrees $0\leq p<n$, and
\bea
\gamma A^n+ dA^{\prime n-1}=0\Longleftrightarrow  \left\{\begin{array}{c}
A^n= \hat a^n+ B^n +\gamma \omega +d \omega',\\ \gamma \hat a^n
=0=\gamma B^n+ dB^{\prime n-1},\\ \left\{\begin{array}{c}\hat  a^n+d\tilde
a^{n-1}+\gamma\omega''=0, \\ \gamma \tilde
a^{n-1}=0,\end{array}\right.\Rightarrow \left\{\begin{array}{c}\hat
a^n=\gamma\omega''',\\ \tilde a^{n-1}=\gamma \omega^4,\end{array}\right.\\
B^{\prime n-1}=\gamma \varpi +d \varpi'\Rightarrow B^n=\gamma \varpi''
+d\varpi,\end{array}\right.
\eea
together with
\bea
\left\{\begin{array}{c} \hat a^n+ B^n
+\gamma \omega +d \omega'=0,\\ \gamma \hat a^n =0=\gamma B^{n}+
dB^{\prime n-1},\\
 B^{\prime n-1}=\gamma \varpi +d \varpi'\Rightarrow B^n=\gamma
\varpi'' +d\varpi\end{array}\right.
\Longrightarrow  \left\{\begin{array}{c} \hat a^n
=dB^{\prime \prime n-1}+\gamma\omega''',\\
\gamma B^{\prime \prime n-1}+dB^{\prime \prime \prime p-2}=0,\\ B^n=\gamma
\varpi'''+ d\varpi^{4},\end{array}\right. \eea
in form degree $n$.

\subsection{Proof of theorem \ref{t1}}\label{s4.4}

That the condition \eqref{4.5} is necessary ($\Longleftarrow$) is
again direct.

\subsubsection{Decomposition according to homogeneity and change of
  variables}

If one decomposes the space of polynomials or formal power series into
monomials of definite homogeneity, the differential $\gamma$ splits
accordingly into a piece that does not change the homogeneity and a
piece that increases the homogeneity by one, $\gamma=\gamma_0
+\gamma_1$.
Consider the change of variables from
$A_\mu,C^a$ and their derivatives to
\bea
y^\alpha\equiv\partial_{(\mu_k}\dots \partial_{\mu_{2}}A^a_{\mu_1)},
\label{4.20}\\
z^\alpha\equiv\partial_{\mu_k}\dots \partial_{\mu_{1}}C^a,\label{4.20ter}\\
C^a,F^{0 a}_\Delta\equiv \partial_{(\mu_{k-1}}\dots
\partial_{\mu_3}F^{0 a}_{\mu_2)\mu_1},\label{4.20bis}
\eea with $F^{0 a}_{\mu\nu} = \partial_{\mu}
A^a_\nu-\partial_\nu A^a_\mu$.
In the new variables,
\bea
\gamma_0=z^\alpha\frac{\partial}{\partial y^\alpha}
\eea
This implies that
\bea
a(y^\alpha,z^\alpha,F^{0a}_\Delta,C^a)=a(0,0,F^{0a}_\Delta,C^a
) +\{\gamma_0,\rho\}a,\\
\rho\, \cdot =\int_0^1\frac{dt}{t} [y^\alpha\frac{\partial}{\partial
z^\alpha}\, \cdot](ty^\alpha,tz^\alpha,F^{0a}_\Delta,C^a),
\eea
Suppose that $\gamma_0 a=0$. It follows that
\bea
\gamma_0 a=0\Longleftrightarrow
a=I+\gamma_0 \omega, \label{prop1}
\eea
for some form valued polynomials $I(x^\mu,dx^\mu,F^{0 a}_\Delta,C^a)$.
Furthermore,
\bea
I+\gamma_0 \omega=0\Longrightarrow I=0.
\label{prop2}
\eea

\subsubsection{Properties of $\gamma_1$ and $\gamma_0$}

Let $\cI^0$ be the algebra of form valued polynomials or formal 
power series that do not depend
on $y^\alpha,z^\alpha$, but only on $C^a,F^{0 a}_\Delta$,
with elements denoted below by $I,J$ and ${\cal P}^0$ be
the algebra of polynomials or formal power series
that depend only on $F^{0a},C^a$, with
elements denoted by $P,Q$. Polynomials which depend only on the ghosts are
denoted by $R$.

Let us introduce the operator
\bea
\sigma=\gamma_1-C^a\delta_a,
\eea
where $\delta_a$ denotes the representation under which the object
transforms. We have
\bea
\sigma A^a_\mu=0,\ \sigma C^a=\half [C,C]^a,\ \sigma^2
=0,\ [\sigma,\partial_\mu]=[\partial_\mu C,\cdot],\ \{\sigma,\gamma_0\}=0.
\eea
The point about $\sigma$ is that when it acts on any expression that depends
only on $A^a_\mu$ and its derivatives, the result does not involve
undifferentiated ghosts. More
precisely, $(\sigma f[A_\mu^a])|_{z=0}=0$. We have $\gamma_0 \sigma
\partial_{\nu_k}\dots \partial_{\nu_2}F^{0a}_{\nu_1\mu}=0$.
It follows that \bea
\sigma \partial_{\nu_k}\dots
\partial_{\nu_2}F^{0a}_{\nu_1\mu}=\gamma_0 \rho\sigma \partial_{\nu_k}\dots
\partial_{\nu_2}F^{0a}_{\nu_1\mu}.
\eea
Symmetrizing over the $\nu$ indices, one gets
\bea
\sigma F^{0a}_{\Delta} = \gamma_0 (\rho\sigma) F^{0a}_{\Delta}.
\eea

\subsubsection{The differential $\gamma^R$}
For later use, let us also establish that if
\bea
\gamma^R J=-[C,F^0_\Delta]^a\frac{\partial J}{\partial
F^{0 a}_\Delta}-\frac{1}{2}[C,C]^a\frac{\partial J}{\partial C^a},
\label{gammar}
\eea
and
\bea
v=(\rho\sigma F^{0a}_{\Delta})
\frac{\partial }{\partial F^{0a}_{\Delta}},
\eea
then
\bea
\gamma_1 J= \gamma_0 (v J)+\gamma^R J.
\label{gool}
\eea

\begin{lemma}\label{l2}
\bea
H(\gamma_1,H(\gamma_0))\simeq H(\gamma^R, \cI^0).
\eea
\end{lemma}
\proof{The lemma means
that
\bea
\left\{\begin{array}{c} \gamma_0 a=0,\\
\gamma_1 a+\gamma_0 b=0,\end{array}\right.\Longleftrightarrow
\left\{\begin{array}{c}a= J+\gamma_0(\ ),\\
\gamma^R J=0, \end{array}\right. \label{coc01}
\eea
and
\bea
\left\{\begin{array}{c} J= \gamma_1 a' +\gamma_0(\ ),\\
\gamma_0  a'=0 ,\end{array}\right. \Longrightarrow   J=\gamma^R J'.
\label{cob01} \eea
Indeed, the result follows directly from \eqref{gool}.
}

Suppose now that the decomposition of the space of polynomials or of
formal power series into monomials of homogeneity $M$ has been
made (see Appendix A for notations and more details).
Then one has the following:
\begin{lemma}\label{l3}
\bea
H(\gamma,\cA)\simeq  \oplus_{M\geq 0} H_M(\gamma^R, \cI^0),\label{e4.57}\\
H(\gamma,\cB)\simeq  \oplus_{M\geq 0} H_M(\gamma^R, \cP^0).
\eea
\end{lemma}
\proof{
Let us first show that every element $[I_M]\in H(\gamma^R, \cI^0) $ can be
completed to a $\gamma$ cocycle. Indeed, let us denote by $I^M$ the
expression obtained by replacing in $I_M$ the variables
$F^{0a}_{\Delta}$ by their non abelian
counterparts
\bea
F^{ a}_\Delta\equiv D_{(\mu_{k-1}}\dots D_{\mu_3}F^{
a}_{\mu_2)\mu_1},\label{nabcounter}
\eea
where $
F^{a}_{\mu\nu} = \partial_{\mu}
A^a_\nu-\partial_\nu A^a_\mu+f^a_{bc}A^b_\mu A^c_\nu$,
and $D_\mu=\partial_\mu+A_\mu^b\delta_b$ and the covariant derivatives of
$F^{a}_{\mu\nu}$ transform in the coadjoint representation. Hence,
$I^M=I_M|$ where the vertical bar
denotes the operation of substitution. Because $\gamma F^a_{\Delta}
=-[C,F_{\Delta}] ^a$, it follows that
\bea
\gamma I^M=(\gamma^R I_M)|=0.\label{completion1}
\eea

Similarily,
\bea
(\gamma^R J_M)|=\gamma J^M. \label{completion2}
\end{eqnarray}
Note that if $I_M,J_M \in \cP^0$, then $I^M\equiv b^M\in
\cB$ and $J^M\equiv \varpi^M\in \cB$.

Suppose $\gamma a^M=0$.
The equations in homogeneity $M$ and $M+1$ imply that $a_M$ is a cocycle of
$H(\gamma_1,H(\gamma_0))$. According to the previous lemma $a_M=I_M+\gamma_0
\eta_{M}$ where $I_M$ is a cocycle of $H(\gamma^R, \cI^0)$.

Suppose that $a^M=\gamma\omega^{M-B}$. In particular, to orders $\leq M$,
we have
\bea
\left\{\begin{array}{ccc}
a_{M} =\gamma_0\omega_M&+&\gamma_1\omega_{M-1}, \\
\gamma_0\omega_{M-1}&+&\gamma_1\omega_{M-2}=0,\\
&\vdots &\\
\gamma_0\omega_{M-B+1}&+&\gamma_1 \omega_{M-B}=0,\\
\gamma_0  \omega_{M-B}=0.
\end{array}\right.
\eea
If $B>
1$, the equations for $\omega_{M-B}$ imply that
$\omega_{M-B}=J_{M-B}+\gamma_0\eta_{M-B}$ with $\gamma^R J_{M-B}=0$. The
redefinition $\omega^{M-B}\rightarrow \omega^{M-B}-\gamma \eta_{M-B}$,
which does
not affect $a^M$ allows to absorb $\eta_{M-B}$. One can then replace
$\omega^{M-B}$ by $\omega^{M-B+1}= \omega^{M-B}-J^{M-B}$, without affecting
$a^M$. This can be done until $B=1$, where one finds $I_M=\gamma^R
J_{M-1}$. This proves that the map $[a^M]\in H(\gamma)\mapsto [I_M]\in
H(\gamma^R, \cI^0)$ is well defined. The map is surjective because as
shown above, every $\gamma^R$ cocycle $I_M$ can be extended to a
$\gamma$ cocycle $I^M$.  It is also injective, because
as also shown above, if $I_M=\gamma^R J_{M-1}$, then $I^M=\gamma
J^{M-1}$, so that $a^M-\gamma (J^{M-1}-\eta_M)=a^{M+1}$ starts at
homogeneity $M+1$.}

\subsubsection{$H(\gamma)$ and split of variables adapted to
  the nonabelian differential $\gamma$}\label{ss4.4.4}

If one is not interested in proving the covariant
Poincar\'e lemma, one can avoid the detour of using the split of
variables adapted to the abelian differential $\gamma_0$ given in
\eqref{4.20}-\eqref{4.20bis} for the characterization of $H(\gamma)$.
One can use instead directly the variables
\bea
Y^\alpha\equiv\partial_{(\mu_k}\dots \partial_{\mu_{2}}A^a_{\mu_1)},
\label{5.1}\\
Z^\alpha\equiv\partial_{(\mu_k}\dots \partial_{\mu_2}D_{\mu_{1})}C^a
\label{5.2},\\
C^a,F^{a}_\Delta\equiv D_{(\mu_{k-1}}\dots
D_{\mu_3}F^{a}_{\mu_2)\mu_1},\label{4.22}
\eea with $F^{a}_{\mu\nu} = \partial_{\mu}
A^a_\nu-\partial_\nu A^a_\mu+f^a_{bc}A_\mu^bA_\nu^c$ adapted to the
non abelian differential $\gamma$, which reads
\bea
\gamma=Z^\alpha\frac{\partial}{\partial
  Y^\alpha}+\gamma^S,\\
\gamma^S=
C^a\delta_a-\half[C,C]^a\frac{\partial}{\partial C^a},\\
\delta_a=-f^c_{ab}F^b_\Delta  \frac{\partial}{\partial
  F^c_\Delta}\label{gammas}
\eea
The usual argument then shows that $H(\gamma,\cA)$
and $H(\gamma,\cB)$ are isomorphic to $H(\gamma^S,\cI)$ respectively
$H(\gamma^S,\cP)$, where
$\cI$ is the algebra of form valued polynomials or formal
power series that do not depend
on $Y^\alpha,Z^\alpha$, but only on $C^a,F^{ a}_\Delta$,
and ${\cal P}$ is
the algebra of polynomials or formal power series
that depend only on $F^{a},C^a$.

\subsubsection{Exterior derivative and contracting homotopy
for the gauge potentials}
Let us introduce the total derivative that does not act on the ghosts,
\bea
\bar \partial_\mu=\partial_\mu-\partial_{(\nu)\mu}
C^a\frac{\partial}{\partial\partial_{(\nu)} C^a},
\eea
and also the associated exterior derivative and contracting homotopy
\cite{Andersonbook},
\bea
\bar d=dx^\mu\bar\partial_{\mu},\\
\bar \rho \omega^p =\int_0^1 dt
\frac{|\lambda|+1}{n-p+|\lambda|+1}\bar \partial_{(\lambda)}
\left(A^a_\mu\left[
\frac{\bar\delta }{\delta \partial_{(\lambda)\nu} A^a_\mu}
\frac{\partial\omega^p}{\partial dx^\nu}\right]
[x,dx,C,tA]\right),\label{explicit}\\
\omega^p[x,dx,C,A]=\omega^p[x,dx,C,0]+\{\bar\rho,\bar d\}\omega^p,\
0\leq p<n.
\eea
where
${\bar\delta }/{\delta \partial_{(\lambda)\nu} A^a_\mu}$ are the higher
order Euler operators with respect to the dependence on $A_{\mu}^a$
only (see \cite{Barnich:2001jy}, Appendix A
for conventions). In the old variables $A_\mu^a, C^a$ and their derivatives,
consider the
split \bea \gamma_1= \gamma^R
+\gamma^{\partial C},\
\gamma^R=[\partial_{(\nu)}A_\mu, C]^a\frac{\partial}{\partial
\partial_{(\nu)}A_\mu^a}-\half [C,C]^a\frac{\partial}{\partial C^a}.
\eea
This is consistent with \eqref{gammar} when acting
on functions that depend only on $F^{0a}_{\Delta},C^a$.
We have
\bea
\{\gamma^R,\bar d\}=0=\{\gamma^R,\bar \rho\}=0.\label{commutationr}
\eea
The first relation is obvious, while the second follows from the
fact that $\gamma^R$ just rotates all the $A^a_\mu$ and their derivatives in
the internal space without changing the derivatives, while $\bar \rho$
only involves the various derivatives and does not act in the internal space.
It can be proved directly using the explicit expression
\eqref{explicit} for $\bar \rho$.

\subsubsection{Core of the proof}

\begin{lemma}\label{l5}
\bea
\left\{\begin{array}{c}\gamma a^{M}=0, \\
da^{M} +\gamma\omega^{M-B}=0  \end{array}\right.\Longrightarrow
\left\{\begin{array}{c}
\gamma^R I_M=0,\\
\bar d I_M+\gamma^R J_{M-1}=0.\end{array}\right.
\label{4.57}
\eea
\end{lemma}
\proof{
In \eqref{4.57}, one can assume without loss of generality that $B=1$.
This can be shown in the same way as in the corresponding part of the proof of
lemma \ref{l3}. We then have
$a_M= I_M+\gamma_0\eta_M$ with $\gamma^R I_M=0$, and
$\omega_{M-1}=J_{M-1}$, with $\gamma^R J_{M-1}=0$. To
order $M$, the l.h.s. of \eqref{4.57} then gives the r.h.s. of
\eqref{4.57}.}
\begin{lemma}\label{l6}
In form degree $<n$,
\bea
\left\{\begin{array}{c}
\gamma^R I_M=0,\\
\bar d I_M+\gamma^R J_{M-1}=0\end{array}\right.\Longleftrightarrow
\left\{\begin{array}{c}I_M=
P_M+\bar d I'_M+\gamma^R J'_{M-1},\\
\gamma^R P_M=0=\gamma^RI'_M.\end{array}\right.\label{4.69}
\eea
\end{lemma}
\proof{That the condition is necessary ($\Longleftarrow$) is
  direct. In order to show that it is sufficient, we are first going
  to show that
\bea
\bar d I_{N a_1\dots a_k}+\gamma^R_k
  J_{N-1 a_1\dots a_k}=0\Longrightarrow \nonumber\\
 I_{N a_1\dots a_k}=P_{N
    a_1\dots a_k}+ \bar d I'_{N a_1\dots a_k} +\gamma^R_k J'_{N-1
    a_1\dots a_k}\label{general},
\eea
where $\gamma^R_k J_{a_1\dots a_k}=-\sum_{l=1}^kC^bf^c_{ba_l}
J_{a_1\dots a_{l-1} c a_{l+1}\dots a_k}+\gamma^R J_{a_1\dots a_k}$.

Indeed, $I_{Na_1\dots a_k}=T_{Na_1\dots a_k}+\tilde I_{Na_1\dots
  a_k}$, where $T_{Na_1\dots a_k}=T_{Na_1\dots a_k}[x,dx,C]$ and 
$\tilde I_{Na_1\dots a_k}[x,dx,C,A=0]=0$, and similarly
$J_{N-1a_1...a_k}=
S_{N-1a_1...a_k}+\tilde J_{N-1a_1...a_k}$. 
 
The l.h.s of condition (\ref{general}) splits into two pieces:
\begin{eqnarray}
\bar d T_{Na_1\dots a_k}+\gamma^R_k S_{N-1a_1...a_k} =0,\label{first}\\
\bar d \tilde I_{Na_1\dots a_k}+\gamma^R_k \tilde J_{N-1a_1...a_k}=0
\label{second}. 
\end{eqnarray}
On forms depending only on $x,dx,C$, $\bar d$ reduces to
$d_x=dx^\mu\frac{\partial}{\partial x^\mu}$. Both $d_x$ 
and the associated contracting homotopy $\rho_x$ of the standard
Poincar\' lemma anticommute with $\gamma^R_k$. From the first equation
(\ref{first}), one then deduces that 
\begin{eqnarray}
T_{Na_1\dots a_k}=R_{Na_1\dots
  a_k}+\gamma^R_k\rho_x S_{N-1a_1...a_k}+\bar d\rho_x T_{Na_1\dots
  a_k}.\label{xc}
\end{eqnarray} 

In form degree zero, the algebraic Poincar\'e lemma of
$\bar{d}$ implies that
\begin{eqnarray}\label{dbargamabar1}
\tilde I_{Na_1...a_k} =  
\bar{\rho}\bar{d}\tilde I_{Na_1...a_k}  = -
\bar{\rho}\gamma^R_k \tilde J_{N-1a_1...a_k} =
 \gamma^R_k \bar{\rho}\tilde J_{N-1a_1...a_k},
\end{eqnarray}
where we have used the l.h.s of the equation (\ref{second})
and the fact that $\gamma^R_k$ anticommutes with $\bar{\rho}$. 

The
action of $\bar{\rho}$ on $\tilde J_{N-1a_1...a_k}$ yields
\begin{eqnarray}
\bar{\rho}\tilde J_{N-1a_1...a_k} = y^\alpha J_{N-2\alpha a_1...a_k} +
\tilde J'_{M-1a_1...a_k},
\end{eqnarray}
Since $\gamma^R_k(y^\alpha \tilde
J_{N-2\alpha a_1...a_k})$ is linear and homogeneous in $y^\alpha$, one
obtains, after  injecting this in the equation (\ref{dbargamabar1}) 
and putting the
$y^\alpha$ to zero, that  
\begin{eqnarray}
\tilde I_{Na_1...a_k}=\gamma^R_k\tilde J'_{N-1a_1...a_k}.
\end{eqnarray}
Together with (\ref{xc}), this gives the r.h.s. of
(\ref{general}) in form degree $0$. 

Suppose now that the result is true in form degrees $< p$ and
that $ I_{N a_1\dots a_k}$ has form degree $p$.
The algebraic Poincar\'e lemma for
$\bar d$ implies that $\tilde I_{N a_1\dots a_k}=\bar \rho \bar d
\tilde I_{N
  a_1\dots a_k}+\bar d\bar
\rho \tilde I_{N a_1\dots a_k}$.
Using \eqref{second}, we get
$\tilde I_{N a_1\dots a_k}=\gamma^R_k\bar \rho \tilde J_{N-1 a_1\dots a_k}
+\bar d \bar \rho
\tilde I_{N a_1\dots a_k}$.
We have $\bar\rho \tilde 
I_{N a_1\dots a_k}=y^\alpha I_{N-1 \alpha a_1\dots a_k}
+\tilde I'_{N a_1\dots a_k}$ and
$\bar \rho \tilde J_{N-1 a_1\dots a_k}=y^\beta 
J_{N-2 \beta a_1\dots a_k}+\tilde J'_{N-1
a_1\dots a_k}$ so that
\bea
\tilde I_{N a_1\dots a_k}- \bar d \tilde I'_{N a_1\dots a_k}-
\gamma^R_k \tilde J'_{N-1
a_1\dots a_k}\nonumber\\
=\bar d (y^\alpha I_{N-1 \alpha a_1\dots a_k})+
\gamma^R_k(y^\beta J_{N-2 \beta a_1\dots a_k}). \label{base}
\eea
The action of $\gamma^R$ only rotates the
$y^\beta$ in the internal space, its action on an expression that is
linear and homogeneous in the $y^\beta$ reproduces an expression that
is linear and homogeneous in $y^\beta$. Supposing that the term in
$y^\alpha I_{N-1 \alpha a_1\dots a_k}$
with the highest number of derivatives on $y^\alpha$ is
$$A^a_{(\mu,\nu_1\dots\nu_m)}I^{(\mu\nu_1\dots\nu_m)}_{N-1 a
  a_1\dots a_k},$$
we get, for the term linear in $y^\alpha$ with the highest number of
derivatives on $y^\alpha$, that
$dx^\sigma A^a_{(\mu,\nu_1\dots\nu_m\sigma)}
I^{(\mu\nu_1\dots\nu_m)}_{N-1 a
  a_1\dots a_k} +\gamma^R_k
(A^a_{(\mu,\nu_1\dots\nu_m\sigma)}J^{(\mu\nu_1\dots\nu_m\sigma)}_{N-2 a
      a_1\dots a_k})=0$. This implies that
$J^{(\mu\nu_1\dots\nu_m\sigma)}_{N-2 a
      a_1\dots a_k}=dx^{(\sigma}J^{\mu\nu_1\dots\nu_m)}_{N-2 a
      a_1\dots a_k}$ and that
$$I^{ (\mu\nu_1\dots\nu_m)}_{N-1 a
  a_1\dots a_k}=dx^{(\nu_m}
I^{(\mu\nu_1\dots\nu_{m-1}))}_{N-1 a
  a_1\dots a_k}-\gamma^R_{k+1} J^{(\mu\nu_1\dots\nu_m)}_{N-2 a
      a_1\dots a_k}.$$
The redefinitions
\bea
y^\alpha I_{N-1 \alpha a_1\dots a_k}&\rightarrow &y^\alpha
I_{N-1 \alpha a_1\dots a_k}-\bar d (
A^a_{(\mu,\nu_1\dots\nu_{m-1})}I^{ (\mu\nu_1\dots\nu_{m-1})}_{N-1 a
  a_1\dots a_k})\nonumber\\&&+\gamma^R_k(
A^a_{(\mu,\nu_1\dots\nu_{m})}J^{(\mu\nu_1\dots\nu_m)}_{N-2 a
      a_1\dots a_k}),\label{redef}\\
y^\beta J_{N-2 \beta a_1\dots a_k}&\rightarrow& y^\beta J_{N-2 \beta
  a_1\dots a_k}
+\bar d (A^a_{(\mu,\nu_1\dots\nu_{m})}J^{(\mu\nu_1\dots\nu_m)}_{N-2 a
      a_1\dots a_k}),
\eea
do not change the equation \eqref{base} and allow to absorb
the term linear in $y^\alpha$ with the highest  number of  derivatives
on $y^\alpha$. These redefinitions can be done until there are no
derivatives on the $y^\alpha$ left, so that
$y^\alpha I_{N-1 \alpha a_1\dots a_k}=A^aI_{N-1 a a_1\dots a_k}$,
$y^\beta J_{N-2 \beta a_1\dots a_k}=A^a J_{N-2 a a_1\dots a_k}$. The
vanishing of the term proportional to $A^a$ in \eqref{base} then implies
that $\bar d I_{N-1 a a_1\dots a_k}+\gamma^R_{k+1} J_{N-2 a a_1\dots
  a_k}=0$. Because this is the l.h.s. of \eqref{general} in form degree
$<p$, we have by induction that  $I_{N-1 a a_1\dots a_k}=P_{N-1 a
  a_1\dots a_k} + \bar d I'_{N-1 a a_1\dots a_k}+\gamma^R_{k+1} J'_{N-2
  a a_1\dots a_k}$. Injecting into \eqref{base} one gets 
\bea
\tilde I_{N a_1\dots a_k}- \bar d \tilde I'_{N a_1\dots a_k}-
\gamma^R_k \tilde J'_{N-1
a_1\dots a_k}=F^a P_{N-1 a
  a_1\dots a_k} \nonumber\\+ \bar d (F^a I'_{N-1 a a_1\dots a_k})
+\gamma^R_k (F^a J'_{N-2
  a a_1\dots a_k}).\label{4.67bis}
\eea
Combining this with \eqref{xc} gives the desired result
\eqref{general}. 

The first line on the right hand side of \eqref{4.69} then follows
as a particular case of \eqref{general}. Applying now $\gamma^R$,
one gets $\gamma^R P_M-\bar d \gamma^R I'_M=0$. Restriction to the
small algebra $\cB$ then implies that $\gamma^R P_M=0=\bar d
\gamma^R I'_M$ because $\bar d (\gamma^R I'_M)|_{\cB} =0$. 

It remains to be proved that $\gamma^R I'_M=0$. Again, we will show
this as the particular case $k=0$ of the fact that $I'_{N a_1\dots
  a_k}$ in \eqref{general} is $\gamma^R_{k}$ closed, 
$
\gamma^R_k I'_{N a_1\dots
  a_k}=0$, if $I_{Na_1\dots a_k}$ is $\gamma^R_k$ closed (and thus
also $\tilde I_{Na_1\dots a_k}$ and $T_{Na_1\dots a_k}$). 

Indeed, in form degree $0$, this holds trivially since
there is no $I'_{N a_1\dots
  a_k}$. Suppose now that in form degree $<p$, \eqref{general} holds
with $ I'_{N a_1\dots
  a_k}$ closed. Using \eqref{commutationr}, it follows that both
$y^\alpha I_{N-1 \alpha a_1\dots a_k}$ and 
$\tilde I'_{N a_1\dots a_k}$ are $\gamma^R_k$ closed. Furthermore, the 
$y^\alpha I_{N-1 \alpha a_1\dots a_k}$ redefined according to 
\eqref{redef} are still $\gamma^R_k$ closed. It follows that $I_{N-1
  aa_1\dots a_k} $ is $\gamma^R_{k+1}$ closed. By induction, it
follows that $I'_{N-1 aa_1\dots a_k}$ is $\gamma^R_{k+1}$
closed. (This also implies that $P_{N-1 aa_1\dots a_k}$ is
$\gamma^R_{k+1}$ closed).This means that $I'_{N a_1\dots a_k}+ F^a
I'_{N-1 aa_1\dots a_k}$ as well as $F^a P_{N-1 aa_1\dots a_k}$ in
\eqref{4.67bis} are $\gamma^R_k$ closed, as was to be shown. 
}

\paragraph{Completing the proof of \eqref{4.5}:} According to
\eqref{completion1}, we can complete $P^M=P_M|$,
$I^{\prime M}=(I'_M)| $ such that $\gamma P^{M}=0=\gamma
I^{\prime M}$, and according to \eqref{completion2} $\gamma^R
J'_{M-1}=\gamma J^{\prime M-1} +O(M+1)$. Hence,
\bea
a^{M}-P^M - d I^{\prime M} -\gamma(\eta_M+A^a\frac{\partial
  I'_M}{\partial C^a}+J^{\prime M-1})=
a^{M+1}.   \label{4.79}
\eea
Because all the individual terms on the left hand side
satisfy the l.h.s. of \eqref{4.5}, so does $a^{M+1}$.
\begin{lemma}\label{l7}
\bea
\left\{\begin{array}{c}b^M+d a^{\prime M-B}+\gamma\omega^{'M-C}=0,\\
\gamma b^M=0=\gamma a^{\prime M-B},\end{array}\right.\Longrightarrow
\left\{\begin{array}{c}P_M+\bar d I'_M+\gamma^R J'_{M-1}=0,\\
\gamma^R P_M=0=\gamma^RI'_M,\end{array}\right.  \label{4.80}
\eea
where $b_M=P_M+\gamma_0\varpi_M$.
\end{lemma}
\proof{Proceeding again as in the proof of lemma \ref{l3}, one can
  assume without loss of generality that $C=B+1\geq
  1$ by suitably modifying $\omega^{\prime M-C}$. If $B\geq 1$, we have by
  assumption at the lowest orders
\bea
\left\{\begin{array}{c}da'_{M-B}+\gamma_1\omega'_{M-B-1}+\gamma_0
\omega'_{M-B}=0,\\
\gamma_0 \omega'_{M-B-1}=0=\gamma_0 a'_{M-B}=\gamma_1a'_{M-B}+\gamma_0
a'_{M-B+1}. \end{array}\right. \eea
We thus have $\omega'_{M-B-1}=J'_{M-B-1}+\gamma_0(\ )$. The $\gamma_0$
exact term can be assumed to be absent by a further modification of
$\omega^{\prime M-B-1}$ by a $\gamma$ exact term that does not affect the
equations. Because $a'_{M-B}=I'_{M-B}+\gamma_0\eta_{M-B}$, we get
\bea
\left\{\begin{array}{c}\bar dI'_{M-B}+\gamma^RJ'_{M-B-1}=0,\\
\gamma^R I'_{M-B}=0. \end{array}\right.
\eea
According to \eqref{4.69}, this implies
\bea
\left\{\begin{array}{c}I'_{M-B}=P^\prime_{ M-B}+\bar d
I''_{M-B}+\gamma^RJ''_{M-B-1},\\ \gamma^RP^\prime_{ M-B}=0=\gamma^R I''_{M-B}.
\end{array}\right.
\eea
As in the reasoning leading to \eqref{4.79}, we get
\bea
 a^{\prime M-B}-P^{\prime M-B} - d I^{\prime\prime M-B}
-\gamma(\eta_{M-B}+A^a\frac{\partial
  I'_{M-B}}{\partial C^a}+J^{\prime\prime M-B-1})\nonumber\\=
a^{\prime M-B+1},\label{4.78}
\eea
with $\gamma  P^{\prime M-B}=0=\gamma I^{\prime \prime M-B}$.
Because $ dP^{\prime M-B} +\gamma(\ )=0$,
we can replace in the l.h.s. of \eqref{4.80}  $a^{\prime M-B}$
by  $a^{\prime M-B+1}$ by
suitably modifying $\omega^{\prime M-C}$. This can be done until $B=0$.
For $B=0$, by the same reasoning as in the beginning of this proof, we
get the r.h.s. of
\eqref{4.80}, with $b_M=P_M+\gamma_0\varpi_M$.
}

\begin{lemma}\label{l8} In form degree $\leq n$, \bea
\left\{\begin{array}{c}P_M+\bar d I'_M+\gamma^R J'_{M-1}=0,\\
\gamma^R P_M=0=\gamma^RI'_M,\end{array}\right.\Longrightarrow
P_M=\gamma^R Q_{M-1}.\label{4.70}
\eea
\end{lemma}
\proof{
By restricting to the small algebra $\cB$, we get
$P_M+\bar d (I'_M|_{\cB})+\gamma^R (J'_{M-1}|_{\cB})=0$. This gives
directly the result because $\bar d (I'_M|_{\cB})=0$.
}

Note that because of the first relation of \eqref{commutationr}, the
map
\bea
\bar d^{\#}:H(\gamma^R,\cI^0)&\longrightarrow& H(\gamma^R,\cI^0),\cr
[I]&\mapsto & [\bar d I],
\eea
is well defined. Lemma \ref{l6} and \ref{l8} (for $p<n$)
can then be summarized by
\begin{corollary}\label{c2}
\bea
H^p(\bar d^{\#},H(\gamma^R,\cI^0))\simeq H^p(\gamma^R,\cP^0),\ p<n.
\eea
\end{corollary}

\paragraph{Completing the proof of \eqref{4.6s}:}
According to \eqref{completion2}, the expression $Q^{M-1}=Q_{M-1}|$
satisfies $\gamma^R Q_{M-1}=\gamma
Q^{M-1}+O(M+1)$ so that $b^M-\gamma (\varpi_M+ Q^{M-1})
=b^{M+1}$. This implies that
$b^{M+1}$ obeys again the l.h.s. of \eqref{4.6s}.

\subsubsection{Convergence in the space of polynomials}

As it stands, the proof of the covariant Poincar\'e lemma is 
valid in the space of formal power
series. In order that they apply to the case of polynomials, one needs
to be sure that if the l.h.s of \eqref{4.5} and \eqref{4.6s} are
polynomials, i.e., if the degrees of homogeneity of all the
elements are bounded from above, then the same holds for the elements
that have been constructed on the r.h.s. of these equations. This can
be done by controling the number of derivatives on the
$A_\mu^a$ and the $C^a$'s. Let
\bea
K=(|\nu|-1)\partial_{(\nu)}C^a\frac{\partial}{\partial \partial_{(\nu)}
C^a} +|\nu|\partial_{(\nu)}A^a_\mu\frac{\partial}{\partial \partial_{(\nu)}
A^a_{\mu}}. \label{Kdegree}
\eea
Suppose that $a^M$ is a polynomial $a^M=a_M+\dots +a_{M+L}$.
It follows that the $K$ degree of $a^M$ is bounded from above by some
$k$. We will say that $a^M$ is of order $k$.
Note that
$\gamma_0$ does not modify the order, while $\gamma_1$ decreases the
order by $1$.
It follows that the
$\gamma_0$ exact term and $J$ in \eqref{coc01}
can be assumed to be of order $k$ as well, while $J'$ in \eqref{cob01}
can be assumed to be of order $k+1$. The important
point is that
$I^M-I_M$ and $\gamma J^M-\gamma^R J_M$ are of order $k-1$
if $I_M$ respectively $\gamma^R J_M$ are of order $k$.

Since $\bar d$ increases the order by $1$,
$I_M$ and $J_{M-1}$ in \eqref{4.57} can be assumed to be of
order $k$, respectively $k+2$, while $I'_M$ can be assumed to be of order
$k-1$. It follows that $J'_{M-1}$ in \eqref{4.69} can be assumed
to be of order $k+1$. This implies that in the recursive construction
\eqref{4.79} of $a^M$, after $M+L+1$ steps, i.e., after all of the
original $a_M+\dots+a_{M+L}$ have been absorbed, the order strictly
decreases at each step. Since the order is bounded from below, the
construction necessarily finishes after a finite number of steps, so
that one stays inside the space of polynomials.

Similarily, if $b^M$ in the l.h.s. \eqref{4.80} is of order $k$, $P_M$, $I'_M$
and $J'_{M-1}$ on the right hand side of \eqref{4.80} can be assumed
to be of order $k$, $k-1$ and $k+1$ respectively. It follows that on
the r.h.s. of \eqref{4.70}, $Q_{M-1}$ can be assumed to be of order
$k+1$. In the recursive construction of $b^M$, once the original $b^M$
has been completely absorbed, the order strictly decreases at
each step, so that the construction again finishes after a finite
number of steps.

\subsubsection{The case of super or graded Lie algebras}

In the case of super or graded Lie algebras, some of the gauge
potentials become fermionic, while some of the ghosts become bosonic.
By taking due care of sign factors and using graded commutators
everywhere, the same proof as above of the covariant Poincar\'e lemma
goes through.

\section{$H(\gamma)$ for $\cG/\cJ$ semisimple}\label{s5}

\subsection{Formulation of a theorem by Hochschild and Serre}

As shown in the digression in subsubsection \ref{ss4.4.4},
$H(\gamma,\cA)\simeq H(\gamma^S,\cI)$, respectively
$H(\gamma,\cB)\simeq H(\gamma^S,\cP)$.
By identifying the ghosts $C^a$ as generators of
$\bigwedge (\cG^*)$, the spaces $\cI$ and $\cP$ can be
identified with $C(\cG,V^\cI)$ respectively $C(\cG,V^\cP)$, the spaces
of cochains with values in the module $V^\cI$ respectively
$V^\cP$. Here, $V^\cI$ is the module of form valued polynomials or
formal power series in the
$F^a_\Delta$, while $V^\cP$ is the module of polynomials or formal
power series in the $F^a$. The differential $\gamma^S$ defined in
\eqref{gammas} can then be identified with the Chevalley-Eilenberg
Lie algebra differential with coeffcients in the module $V^\cI$
respectively $V^\cP$. The module $V^\cP$ decomposes into the
direct sum of finitedimensional modules $V^\cP_M$ of monomials of
homogeneity $M$ in the $F^a$. The module $V_\cI$ decomposes
into the direct sum of modules $\Omega(M)\otimes V^\cI_M$ of
form valued monomials of homogeneity $M$ in the $F^a_\Delta$.
The spacetime forms can be factorized because the representation
does not act on them, and the module $V^\cI_M$ is finitedimensional.

As mentioned in the introduction, it is at this stage that, for
reductive Lie algebras, one can use
standard results on Lie algebra cohomology with coefficients in a
finitedimensional module (see e.g. \cite{Greub}).
But even for non reductive Lie
algebras, there exist some general results. We will now review one
of these results due to Hochschild and Serre \cite{HochschildSerre}.
In order to be self-contained, a simple proof of their theorem
in ``ghost'' language is given.

\begin{theorem}{ Let ${\cal G}$ be a real Lie algebra and
${\cal J}$ an ideal of
${\cal G}$ such that ${\cal G}/{\cal J}$ is semi-simple. Let
$V$ be a finite dimensional ${\cal G}$-module. Then the following
isomorphism holds
\begin{equation} H({\cal G}, V) \simeq H({\cal G}/{\cal J},
  \mathbb{R})
\otimes
H^{{\cal G}}({\cal J}, V)\end{equation}}where $H^{{\cal G}}$
means the ${\cal G}$-invariant cohomology space.
\end{theorem}

\subsection{Proof}

The above hypothesis implies that there is a semisimple
subalgebra ${\cal K}$ of
${\cal G}$ isomorphic to
${\cal G}/{\cal J}$ such that \begin{equation}{\cal G} = {\cal K} \ltimes {\cal
J}.\end{equation}
 Let $\{e_A, h_\alpha\}, (A=1,...,p),
(\alpha=1,...,q)$ denote a basis of
${\cal G}$, among which the $e_A$'s form a basis of ${\cal
K}$ and the
$h_\alpha$'s a basis of ${\cal J}$: the fundamental brackets are given by
\begin{equation}
[e_A, e_B] = f_{AB}^{\,\,\,\,\,C}\,e_C,\quad [h_\alpha, h_\beta]
= f_{\alpha \beta}^{\,\,\,\,\,\gamma}\,h_\gamma,\quad [e_A,h_\beta] =
f_{A \beta}^{\,\,\,\,\,\gamma}\,h_\gamma.
\end{equation}
If $C^a\equiv(\eta^A,C^\alpha)$, the coboundary operator
$\gamma^S$ can be cast into the form
\begin{align}\gamma^S&=\eta^{A}\,\rho(e_A) +
\eta^{A}\,\rho_C(e_A) - \frac{1}{2}\,\eta^{A}\,\eta^B\,f_{AB}^{\,\,\,\,\,C}\,
\,\frac{\partial}{\partial \eta^C} +\nonumber\\ &+
C^{\alpha}\,\rho(h_\alpha) -
\frac{1}{2}\,C^{\alpha}\,C^\beta\,f_{\alpha \beta}^{\,\,\,\,\,\gamma}\,
\frac{\partial}{\partial C^\gamma},
\end{align}
Here, $\rho_C(e_A)$
is the extension to $\bigwedge(C)$ of the coadjoint
representation of the semi-simple ${\cal K}$,
\bea
\rho_C(e_A) = -\,
f_{A\beta}^{\,\,\,\,\,\gamma}\,C^\beta\,\frac{\partial}{\partial
C^\gamma},
\eea
while $\rho$ denotes
the representation of ${\cal G}$ in $V$.
Let
\begin{equation} N_\eta =
\eta^{A}\,\frac{\partial}{\partial \eta^{A}}\quad\quad N_C =
C^\alpha\,\frac{\partial}{\partial C^\alpha}\end{equation} be the
counting operators for the $\eta$'s and $C$'s and the associated
gradings $gh_\eta$ and $gh_C$ on $V\otimes
\bigwedge (C,\eta)$. According to the $gh_C$-grading, $\gamma^S$ is the sum
\begin{equation} \gamma^S= \gamma^S_0 + \gamma^S_1,\quad (\gamma^S)^2 =
(\gamma^S_0)^2 = (\gamma^S_1)^2 = 0,\quad \{\gamma^S_0, \gamma^S_1\}
= 0\end{equation}
with $\gamma^S_1$ explicitly given by
\begin{equation} \gamma^S_1 = C^{\alpha}\,\rho(h_\alpha) -
\frac{1}{2}\,C^{\alpha}C^\beta\,\,f_{\alpha \beta}^{\,\,\,\,\,\gamma}\,
\frac{\partial}{\partial C^\gamma} \end{equation} and which obey
\begin{equation} [N_C, \gamma^S_0] = 0\quad\quad [N_C, \gamma^S_1] =
\gamma^S_1,
\end{equation}
which means that $\gamma^S_0$ conserves the number of
$C$'s while $\gamma^S_1$ increases this number by one. At this stage,
$\gamma^S_0$ can already be identified with the coboundary operator of the Lie
algebra cohomology of the semi-simple sub-algebra ${\cal K}$ with
coefficients in the
${\cal K}$-module $V\otimes \bigwedge(C)$, the corresponding representation
being defined as $\rho^T(e_A) = \rho(e_A)\otimes {\mathbb I}_C +
\mathbb{I}_V\otimes\rho_C(e_A)$. An element
$ a\in V\otimes
\bigwedge (C,\eta)$  of total ghost number $g$ can
be decomposed according to its $gh_C$ components,
\begin{equation} a
= a_0 + a_1 + ... + a_g, \quad gh_C\,a_k = k.
\end{equation}
The cocycle condition $\gamma^S a = 0$ generates the
following tower of equations
\begin{align}
\gamma^S_0\,a_0 &= 0, \label{un}\\ \gamma^S_1\,a_0 +
\gamma^S_0\,a_1 &= 0, \label{deux}\\ \gamma^S_1\,a_1 + \gamma^S_0\,a_2 &=
0,\label{trois}\\ &\vdots\nonumber\\
\gamma^S_1\,a_g &= 0,
\label{quu}\end{align} and the coboundary condition reads
\begin{align}a_0 &=
\gamma^S_0\,\omega_0,\\ a_1 &= \gamma^S_1\,\omega_0 + \gamma^S_0\,\omega_1, \\
\vdots&\nonumber
\\ a_g
&=
\gamma^S_1\,\omega_{g-1} .
\end{align}
The above mentioned results on reductive Lie algebra cohomology imply
that the general solution of equation
\eqref{un} can be written as
\begin{align}& a_0 =
v_0^j\,\Theta_j +
\gamma^S_0\,\omega_0 \label{solution}\\ &\rho^T(e_A)\,v_0^j =
0\end{align} where the $\Theta_j(\eta)$'s form a basis of the
cohomology $H({\cal K}, \mathbb{R})$, which is generated by the
primitive elements. Furthermore, all ${\cal K}$-invariant polynomials
$v^j$ obeying
$v^j\,\Theta_j + \gamma^S_0\,\omega = 0$ for some $\omega$,
have to vanish, $v^j =
0$.

The term
$\gamma^S_0\,\omega_0$ can be absorbed by subtracting
$\gamma^S\omega_0$ from $a$ and modifying $a_1$ appropriately.
Injecting then \eqref{solution} in \eqref{deux}, one gets,
since $\gamma^S_1$
doesn't act on the $\eta$'s,
\begin{equation}
(\gamma^S_1\,v_0^j)\,\Theta_j +
\gamma^S_0\,a_1 = 0.\end{equation}
Now, from $[\rho^T(e_A),
\gamma^S_1] = 0,$ one sees that $\gamma^S_1 v_0^j \in V\otimes \bigwedge(C)$
is still invariant under
$\rho^T(e_A)$.
Accordingly, one must have
\begin{equation}
\gamma^S_1\,v_0^j = 0\mbox{\quad
and}\quad
\gamma^S_0\,a_1 = 0.\label{deltazeroa1}\end{equation}
Again, the general solution of the last
equation
\eqref{deltazeroa1} is
\begin{equation}
a_1 = v_1^j\,\Theta_j +
\gamma^S_0\,\omega_1\end{equation} with $\rho^T(e_A)\,v_1^j = 0$;
subtraction of $\gamma^S\omega_1$ and injection in
equation \eqref{trois} gives
\begin{equation} (\gamma^S_1\,a_1^j)\,\Theta_j + \gamma^S_0\,a_2 =
0,\end{equation} implying
\begin{equation}
\gamma^S_1\,v_1^j = 0\mbox{\quad and}\quad \gamma^S_0\,a_2 =
0.\end{equation}
The same procedure can be repeated until
\eqref{quu}.

Every $\gamma^S$-cocycle is thus of the form
\begin{equation} a = \sum_{k=0}^g\,v_k^j\,\Theta_j +
\gamma^S\,\omega,\label{hscoc}
\end{equation}
with
\begin{align}
\rho^T(e_A)\,v_k^j &= 0 \Longrightarrow
v_k^j\in [V\otimes\bigwedge(C)]^{\cal K}\\
\gamma^S_1\,v_k^j &= 0.
\end{align}
Let us now analyze the coboundary condition. To order $0$, we find
\begin{equation}
v^j_0=0\mbox{\quad and}\quad \gamma_0^S\omega_0=0.
\end{equation}
The last equation implies $\omega_0=w^j_0\Theta_j+\gamma^S_0(\ )$. The
$\gamma^S_0$ exact term can be absorbed by subtracting the
corresponding $\gamma^S$ exact term from $\omega$. To order $1$, we
then find
\bea
v^j_1=\gamma^S_1 w^j_0\mbox{\quad and}\quad \gamma_0^S\omega_1=0.
\eea
Going on in the same way gives
\bea
v^j_k=\gamma^S_1w^j_{k-1},\ k=1,\dots,g.\label{hscob}
\eea
In other words,
\bea
H(\cG,V)\simeq H(\cG/\cJ,\mathbb{R})\otimes
H(\gamma^S_1,(V\otimes\bigwedge(C))^\cK).
\eea
From $\{\gamma^S_1, \frac{\partial}{\partial
C^\alpha}\} = \rho^T(h_\alpha)$, it follows that the elements
$[v^j_k]$ of the second space
are invariant under the action of
${\cal J}$,
\begin{equation}
\rho^T(h_\alpha)\,v_k^j =
 \gamma^S_1\,\frac{\partial}{\partial
C^\alpha}\,v_k^j\,\Longrightarrow
(\rho^T(h_\alpha))^{\#}[v_k^j]=0,
\end{equation} where $\rho^T(h_\alpha) = \rho(h_\alpha)\otimes
{\mathbb I}_C+{\mathbb I}_V\otimes \rho_C(h_\alpha)$. Hence,
\bea
H(\gamma^S_1,(V\otimes\bigwedge(C))^\cK)=H^\cG(\cJ,V),
\eea
as required.

\subsection{Explicit computation of
$H(\gamma,\cB)$ for ${\cal G} = iso(3)$ or $iso(2,1)$}

\subsubsection{Applicability of the theorem}

As a concrete application, we consider the case where $\cG=iso(3)$,
the 3 dimensional Euclidian algebra, or $\cG=iso(2,1)$, the 3
dimensional Poincar\'e algebra. Both of these Lie algebras
fulfill the hypothesis of the
Hochschild-Serre theorem with ${\cal J}$ being the abelian translation
algebra.

Denoting by $\{h_a=P_a, e_a =J_a\}$ a basis of ${\cal G}$ where
$P_a$ represent the translation generators and $J_a$ represent the rotation
(resp. Lorentz) generators, their brackets can be written as
\begin{equation} [P_a, P_b] = 0,\quad [J_a, J_b] =
\epsilon_{abc}\,J^c ,\quad [J_a, P_b] = \epsilon_{abc}\,P^c.\end{equation}
The indices are lowered or raised with the Killing metric $g_{ab}$ of the
semi-simple subalgebra ${\cal K} = so(3)$ or $so(2,1)$.

In the so called
universal algebra, (see
\cite{Dubois-Violette:1985jb,Barnich:2000zw} for more details) the space of
polynomials in the $F^a$, the abelian curvature $2$-form associated
to the translations, and the $G^a$, the non abelian curvature $2$-form
associated to the rotations/boosts, can be identified with the
module $V = S({\cal G^*})$ transforming under the
the extension of the coadjoint representation, so that
\bea
H(\gamma,\cB))\simeq H(\gamma^R,\cP)\equiv H(\cG,S(\cG^*)).
\eea

The coboundary operator $\gamma^R$ acts on $V
\otimes\bigwedge(C,\eta)$ through
\begin{equation}
\gamma^R =
\eta^{a}\,\epsilon_{abc}\,[F^c\,\frac{\partial}{\partial F_b} +
G^c\,\frac{\partial}{\partial G_b} -
\frac{1}{2}\,\eta^b\,\frac{\partial}{\partial \eta_c} -
C^b\,\frac{\partial}{\partial C_c}\,] +
C^{a}\,\epsilon_{abc}\,G^c\,\frac{\partial}{\partial
F_b}.
\end{equation}

As mentioned above, the Lie algebra cohomology $H({\cal K},
\mathbb{R})$ is generated by particular ghost polynomials $\Theta_i(\eta)$
representing the primitive elements which are in
one to one correspondence with the independent Casimir operators. In
the particular cases considered here, there is but one primitive element
given by
\begin{equation} \theta_1 =
\frac{1}{3!}\,\epsilon_{abc}\,\eta^{a}\,\eta^b\,\eta^c =
(-)^\sigma\,\hat\eta^3, \end{equation}
where $\sigma=0,1$ for the Euclidean respectively Minkowskian
case. The elements of the set
$\{ 1, \theta_1\}$ provide a basis of this cohomology.

\subsubsection{Invariants, Cocycles, Coboundaries}

\paragraph{Order zero}

In $gh_C=0$, the invariant space $V^{\cal K}$ is generated by the following
quadratic invariants
\begin{equation} f_1 = g_{ab}\,G^a\,G^b,\quad f_2 = g_{ab}\,F^a\,F^b,\quad
f_3 = g_{ab}\,F^a\,G^b.\end{equation} An element $a_0 \in V^{\cal K}$ is a
polynomial in the 3 variables
\begin{equation} a_0 = Q( f_1, f_2, f_3).\end{equation} To fulfill the
cocycle condition
$\gamma^R_1 a_0 = 0$, $Q$ has to obey
\begin{equation} \epsilon_{abc}\,G^c\,\frac{\partial}{\partial
F_b}\,Q = 0 = \epsilon_{abc}\,G^c\,[ 2\,F^b\,\frac{\partial}{\partial
f_2} + G^b\,\frac{\partial}{\partial
f_3}\,]\,Q, \end{equation} which implies \begin{equation}
\frac{\partial}{\partial f_2}\,Q = 0.\end{equation} The
$\gamma^R_1$-cocycles of $gh_C=0$ are thus of the form
\begin{equation} a_0 = Q(f_1, f_3)\end{equation} Using the following
decomposition
\begin{equation} Q(f_1, f_2, f_3) = Q(f_1,0,f_3) + f_2\,\tilde
Q(f_1,f_2,f_3),\end{equation}
the coboundaries of $gh_C=1$ are given by
\begin{equation}
t_1 = \gamma^R_1\,[f_2\,\tilde Q(f_1,f_2,f_3)] =
2\,C^{a}\epsilon_{abc}\,G^c\,F^b\,\frac{\partial}{\partial f_2}\,[f_2\,\tilde
Q(f_1,f_2,f_3)].\label{cob1}
\end{equation}

\paragraph{Order one}

In $gh_C=1$, the elements of $[V\otimes \bigwedge(C)]^{\cal K}$
can be written as
\begin{equation} \omega_1 = C^b\,\omega_b\,\end{equation} where the
$\omega_b \in S({\cal
G}^*)$ transform under ${\cal K}$ as the components of a vector. They are of
the form
\begin{equation} C^b\,\omega_b = C^b[G_b\,Q(f_k) + F_b\,R(f_k) +
\epsilon_{bcd}\,G^c\,F^d\,S(f_k)].\end{equation}
According to
\eqref{cob1}, the last term is
$\gamma^R_1$-exact.
For the other terms, the cocycle condition $\gamma^R_1 a_1 = 0$
implies
\begin{equation}
C^{a}C^b\epsilon_{amn}G^n\,[2G_bF^m\frac{\partial
Q}{\partial f_2} + \delta_b^m R + 2F_bF^m\frac{\partial R}{\partial f_2}] =
0,
\end{equation}
and imposes
\begin{equation}
R = 0\mbox{\quad and}\quad
\frac{\partial Q}{\partial f_2} = 0.\end{equation}
Hence, the non-trivial $gh_C =
1$ cocycles are given by
\begin{equation}
a_1 = C^b\,G_b\,Q(f_1,f_3).
\end{equation}
The $gh_C=2$ coboundaries are given by
\begin{equation}
t_2 =\gamma^R_1\,C^b\,[G_b\,f_2\,\tilde Q(f_k) +
F_b\,R(f_k)],
\end{equation}
or equivalently by
\begin{align} t_2 &= [\,(GC^2)\,f_3 - (FC^2)\,f_1]\,\frac{\partial
f_2\tilde Q}{\partial f_2}\nonumber\\ &+ (GC^2)\,R + [(GC^2)\,f_2
-(FC^2)\,f_3\,]\frac{\partial R}{\partial f_2}\label{cobo2}
\end{align}
due to the identities
\begin{align}
2\,(C^{a}G_a)\,(C^b\epsilon_{bcd}G^cF^d) &= (GC^2)\, f_3 -
(FC^2)\,f_1 \\ 2\,(C^{a}F_a)\,(C^b\epsilon_{bcd}G^cF^d) &=
(GC^2)\,f_2 - (FC^2)\, f_3\end{align} in which $(FC^2) =
C^{a}C^b\epsilon_{abc}F^{c}$ and $(GC^2)
=C^{a}C^b\epsilon_{abc}G^{c}$.

One deduces from \eqref{cobo2} that, for all integers $L,M,N$, the following
equalities between $\gamma^R_1$ equivalences classes hold:
\begin{align} [(GC^2)\,f_1^L\,f_2^M\,f_3^{N+1}] &=
[(FC^2)\,f_1^{L+1}\,f_2^{M}\,f_3^{N}], \\ [(GC^2)\,f_1^L\,f_2^M\,f_3^N] &=
[\frac{M}{M+1}\,(FC^2)\,f_1^{L}\,f_2^{M-1}\,f_3^{N+1}],
\end{align}
from which one infers that the elements of the form $(GC^2)\,U(f_k)$
are equivalent to
elements of the form $(FC^2)\,V(f_k)$
and furthermore that all monomials
of the form
$(FC^2)\,f_1^L\,f_2^M\,f_3^N $ with $L>M$ are coboundaries,
while those which have $L\leq M$ can be replaced by monomials not
containing $f_1$ according to
\begin{equation}[(FC^2)\,f_1^L\,f_2^M\,f_3^N]=
[(FC^2)\,\frac{(M-L+1)}{M+1}\,f_2^{M-L}\,f_3^{N+2L}].
\end{equation}

\paragraph{Order two}

The $gh_C=2$, elements of $[V\otimes \bigwedge(C)]^{\cal K}$ can be
written as
\begin{equation} \omega_2 = C^{a}\,C^b\,\omega_{ab} .
\end{equation}
The most general element is of the type
\begin{equation} \omega_2 = C^{a}\,C^b\,\epsilon_{abc}\,[G^c\,U(f_k) +
F^c\,V(f_k) +
\epsilon^{cmn}\,G_mF_n\,W(f_k)],
\end{equation}
but, according to our
preceeding results, through the addition of an appropriate coboundary, we
can remove the
$U$-part and suppose $V$ not depending on $f_1$.
The cocycle condition
$\gamma^R_1 a_2 = 0$ then reads
\begin{equation}
C^d\,C^{a}\,C^b\,\epsilon_{def}\,G^f\,\epsilon_{abc}\,\,\frac{\partial
}{\partial F_e}[F^c\,V(f_2,f_3) + \epsilon^{cmn}\,G_mF_n\,W(f_k)]
= 0.\end{equation}
It does not further restrict $V$ but requires $W=0$. The non trivial
$gh_C= 2$ cocycles are thus given by
\begin{equation} a_2 = (FC^2)\,V(f_2, f_3).
\end{equation} In order to
characterize the $gh_C = 3$ coboundaries, we use the following
identity
\begin{align}
\gamma^R_1\,(F\times G)C^2\,f_1^L\,f_2^M\,f_3^N &=
C^{1}\,C^2\,C^3\,(\,4(M+1)\,f_1^{L+1}\,f_2^M\,f_3^N \nonumber\\ &-
4M\,f_1^L\,f_2^{M-1}\,f_3^{N+2})\,\end{align} from which we infer that all
monomials of the form
$C^1C^2C^3\,f_1^L\,f_2^M\,f_3^N$ for $L> M$ are coboundaries, while
those with
$L\leq M$ are equivalent to monomials involving powers of $f_2$ and $f_3$
only.

\paragraph{Order three}

The $gh_C = 3$ invariants are of the form \begin{equation} \omega_3
= C^{1}\,C^2\,C^3\,Q(f_1,f_2,f_3).\end{equation} All of them are cocycles
since ${\cal J}$ is of dimension 3, but only those of the form
\begin{equation} a_3
= C^{1}\,C^2\,C^3\,Q(f_2,f_3)
\end{equation} are non-trivial.

\paragraph{Summary}

The non-trivial cocycles of $H(\gamma^R_1,[V\otimes
\bigwedge(C)]^{\cal K})$ are summarized in table 1, where
$\hat C^3=C^1C^2C^3$, $CG=C^aG_a$.
They provide a basis of
$H(\gamma^R_1,[V\otimes \bigwedge(C)]^{\cal K})$ as a vector space.

\begin{table}[tbp]
\centerline{
\begin{tabular}{|c|c|}
\hline
 $gh_C$ & $H(\gamma^R_1,[V\otimes
\bigwedge(C)]^{\cal K})$ \\
\hline\hline
$0$ & $Q(f_1, f_3)$ \\
\hline
$1$ & $C\,G\,R(f_1,f_3)$ \\ \hline
$2$ & $FC^2\,S(f_2, f_3)$ \\ \hline
$3$ & $\hat C^3S(f_2,f_3)$\\
\hline
\end{tabular}\\[1ex]
}
\hspace*{5.5cm}
\emph {Table 1}
\end{table}

The associated basis of $H(\gamma,\cB)\simeq H(\gamma_1^R,\cP)$ is
given by
\begin{align}
&\{ Q_0(f_1,
f_3),\,\, CG \, R_0(f_1,f_3),\,\,
FC^2\, S_0(f_2,
f_3),\,\, \hat C^3\, T_0(f_2,f_3),\nonumber\\
&\hat \eta^3\, Q_1(f_1, f_3),\,\,
\hat \eta^3\,CG\,R_1(f_1,f_3),\,\,\hat\eta^3\, FC^2\, S_1(f_2,
f_3),\,\, \hat\eta^3\, \hat C^3\, T_1(f_2,f_3)\}.
\label{Bw}\end{align}

\section{$H(\gamma|d)$ for $\cG/\cJ$
  semisimple and $\cJ$ abelian}\label{s6}

Let $\cK$ be a semisimple Lie algebra and $\cG=\cK\ltimes\cJ$
with $\cJ$ an abelian ideal.  This means that, with respect to
section \ref{s5}, the additional assumption $[h_\alpha,h_\beta]=0$
holds. In other words, the only possibly
non vanishing structure constants are given
by $f^C_{AB}$ and $f^\gamma_{A\beta}$.

\subsection{$H(\gamma|d,\cB)$}

\subsubsection{General results}
Let $B^A$ and $\eta^A$ the gauge
field $1$-forms and ghosts associated to $\cK$ and $A^\alpha$ and
$C^\alpha$ the gauge fields $1$-forms and ghosts associated to $\cJ$.
The curvature $2$ form decomposes as $G^A=dB^A+\frac{1}{2}[B,B]^A$ and
$F^\alpha=dA^\alpha+[B,A]^\alpha$.
Let us consider the algebra ${\cal B}$ using the variables
$C^\alpha$, $DC^\alpha=dC^\alpha+[B,C]^\alpha$, $A^\alpha$,
$F^\alpha$, $B^A$,$G^A$, $\eta^A$,
$D\eta^A=d\eta^A+[B,\eta]^A$. Applying the results of section \ref{s3},
we have
\bea
d=[\lambda,\gamma].\label{6.11}
\eea
As in section \ref{s3}, if $\gamma b=0$, one has
\bea
db+\gamma\lambda b=0,\label{domega}
\eea
and
\bea
d\lambda b+\gamma\frac{1}{2}\lambda^2 b=\tau
b,\label{dlomega}
\eea
with
\bea
\tau=\frac{1}{2}[d,\lambda]=F^\alpha\frac{\partial}{\partial C^\alpha}+
G^A\frac{\partial}{\partial
\eta^A},\\
\tau^2=0,\ \{\tau,\gamma\}=0.
\eea
Furthermore, if
\bea
\sigma=C^\alpha\frac{\partial}{\partial F^\alpha},
\eea
\bea
\sigma^2=0,\ \{\tau,\sigma\}=N_{C,F},\label{tausigma}
\\
\{\sigma,\gamma\}=0,\label{taugamma}
\eea
It is in order for this last relation to hold that one needs the
assumption that $\cJ$ is abelian. Indeed, in this case,
because
\bea
\gamma=-DC^\alpha\frac{\partial}{\partial A^\alpha}-D\eta^A
\frac{\partial}{\partial B^A}
+([F,\eta]+[G,C])^\alpha\frac{\partial}{\partial F^\alpha}+[G,\eta]^A
\frac{\partial}{\partial G^A}\nonumber\\-[\eta,C]^\alpha
\frac{\partial}{\partial C^\alpha}-\half[\eta,\eta]^A
\frac{\partial}{\partial \eta^A},
\eea
the absence of the term $[F,C]^\alpha\partial/\partial F^\alpha$
guarantess that \eqref{taugamma} holds.

According to \eqref{hscoc}, we can assume
$b = v^j\Theta_j$, where $v^j=v^j(F,G,C)$ with
\bea
\rho^T(e_A)v^j=0=\gamma v^j,\\
\rho^T(e_A)=-f^\gamma_{A\beta}F^\beta\frac{\partial}{\partial
  F^\gamma}-f^C_{AB}G^B\frac{\partial}
{\partial G^C} -f^\gamma_{A\beta}C^\beta\frac{\partial}{\partial C^\gamma},\\
\gamma v^j=-[C,G]^\alpha\frac{\partial}{\partial
F^\alpha} v^j.
\eea
Applying (\ref{domega}) and \eqref{dlomega}, we get
\bea
d v^j+\gamma\lambda v^j=0,\\
d\lambda v^j +\gamma\frac{1}{2}\lambda^2 v^j=\tau
v^j.
\eea
Furthermore, because $\cK$ is semisimple, there exist
$\hat\Theta_j$ and $\hat{\hat{\Theta}}_j$ such that
\bea
d\Theta_j+\gamma \hat\Theta_j=0,\\
d\hat\Theta_j+\gamma\hat{\hat{\Theta}}_j=0.
\eea
It follows that
\bea
\gamma(v^j\Theta_j)=0,\\
d(v^j\Theta_j)+\gamma((\lambda
v^j)\Theta_j+v^j\hat\Theta_j)=0,\label{6.26}\\
d((\lambda v^j)\Theta_j+v^j\hat\Theta_j)+\gamma((\half\lambda^2
v^j)\Theta_j+(\lambda v^j)\hat\Theta_j+v^j\hat{\hat{\Theta}}_j)=(\tau
v^j)\Theta_j.\label{6.27}
\eea
The necessary and sufficient condition
that $v^j\Theta_j$ ``can be
lifted twice'', i.e; that $[v^j\Theta_j]\in {\rm Ker}\ d_1$, with
\bea
d_1: H(d,H(\gamma,\cB)) &\longrightarrow&
H(d,H(\gamma,\cB)), \nonumber\cr
[v^j\Theta_j] &\mapsto& d_1[v^j\Theta_j]=[d((\lambda
v^j)\Theta_j+v^j\hat\Theta_j)],
\eea
is
\bea
d((\lambda
v^j)\Theta_j+v^j\hat\Theta_j)=d b'+\gamma(\ ),
\eea
with $\gamma b'=0$. Because $db'+\gamma(\ )=0$, it follows by using
\eqref{6.27} that this
necessary and sufficient condition is
\bea
(\tau v^j)\Theta_j+\gamma(\ )=0.
\eea
Because $\tau$ commutes with $\rho^T(e_a)$ and anticommutes with
$\gamma$, it follows from \eqref{hscob} that the condition reduces to
\bea
(\tau v^j)+\gamma w^j=0,\ \rho^T(e_A)w^j=0,
\eea
for some $w^j$.
Let us decompose $v^j$ as a sum of terms of definite
$N_{F,C}$ degree $k$, $v^j = v^j_0 + \sum_{k=1}v^j_k$.
Using
(\ref{tausigma}), this can be rewritten as
\bea
v^j =  v^j_0+ \sum_{k=1}\,( \sigma t^j_{k} +
\tau s^j_{k}),
\eea
where $t^j_{k}=1/k\,\tau v^j_k$ and $s^j_{k}=1/k\, \sigma v^j_k$.
This decomposition is direct and induces a well defined decomposition in
cohomology because $\gamma$ and $\tau$, respectively $\sigma$,
anticommute.  Furthermore,
\bea
\tau v^j_0 = 0,\ \tau (\tau s^j_k) = 0,\\
\tau \sigma t^j_k +\gamma w^j_k=0\Longrightarrow
\tau v^j_k+\gamma w^j_k=0\Longrightarrow
\sigma t^j_k=\gamma \frac{1}{k}\sigma w^j_{k} = 0.
\eea
This implies for the decomposition $H(\gamma,\cB)= E_2\oplus d_1
F_1\oplus F_1$, with ${\rm Ker}\ d_1=E_2\oplus d_1
F_1$, that
\bea
{\rm Ker}\ d_1 = \{v^j_0\Theta_j + \sum_{k=1}\,
[\tau s^j_k]\,\Theta_j\}\\ d_1 F_1=
\{ \sum_{k=1}\,[\tau s^j_k]\,\Theta_j\}\\ F_1 = \{
\sum_{k=1} [\sigma t^j_k]\,\Theta_j\}\\
 E_2 = \{v^j_0\,\Theta_j\}.
\eea
Here, $[\tau s^j_k]$ and $[\sigma t^j_k]$ denote equivalence classes
of $\rho^T(e_A)$ invariant cocycles that are $\tau$, respectively
$\sigma$ exact, up to coboundaries of $\rho^T(e_a)$ invariant
elements that are also $\tau$, respectively
$\sigma$ exact.

Let
\bea
\lambda^{\#}: F_1&\longrightarrow& H(\gamma|d,B),\cr
[\sigma t^j_k]\Theta_j &\mapsto&
[(\lambda\sigma
t^j_k)\Theta_j+\sigma t^j_k\hat\Theta_j].
\eea
That the map is well defined follows from \eqref{6.26} and $\lambda
\gamma(\sigma w^j_k)\Theta_j+\gamma(\sigma w^j_k)\hat\Theta_j=d(\sigma
w^j_k\Theta_j+\gamma(\lambda\sigma w^j_k\Theta_j+\sigma
w^j_k\hat\Theta_j)$ due to \eqref{6.11}.

Let $\cB_\cK$ be the restriction of $\cB$ to the generators associated
to $\cK$. Because $E_2\simeq H(\gamma,\cB_\cK)$, we have
\bea
H(\gamma,\cB_\cG)\simeq H(\gamma,\cB_\cK)\oplus d_1F_1\oplus F_1.
\eea
Furthermore, the general analysis of the exact triangle associated to
the descent equations \cite{Dubois-Violette:1985jb} (see also
e.g. \cite{Henneaux:1998rp}) implies that
\bea
H(\gamma|d,\cB_\cG)\simeq H(\gamma|d,\cB_\cK)\oplus \lambda^{\#}
F_1\oplus F_1.
\eea
This solves the problem because the classification of
$H(\gamma|d,\cB_\cK)$  and the associated decomposition of
$H(\gamma,\cB_\cK)$ for semisimple $\cK$ has been completely solved
\cite{Dubois-Violette:1985jb} (see e.g. \cite{Barnich:2000zw} for a
review).

\subsubsection{Application to $\cG=iso(3)$ or $iso(2,1)$}

By applying the analysis of the previous subsubsection to the
particular case of
$iso(3)$, respectively $iso(2,1)$, with $H(\gamma,\cB)$ given by
\eqref{Bw}, it follows that
\begin{align}
F_1 &=
\{GC\,R_0(f_1,f_3),\,\,\,\hat C^3\,T_0(f_2,f_3),\,\, \hat
\eta^3\,GC\,R_1(f_1,f_3),\,\,\hat C^3\hat\eta^3\,T_1(f_2,f_3)\},\\
d_1 F_1&=\{f_3\,\tilde
Q_0(f_1,f_3),\,\,FC^2\,S_0(f_2,f_3),\,\,\hat \eta^3\,f_3\,\tilde
Q_1(f_2,f_3),\,\,\hat\eta^3FC^2\,S_1(f_2,f_3)\},\\
E_2&=\{ Q_0(f_1), \hat\eta^3\,Q_1(f_1)\}.
\end{align}
Furthermore, the general analysis of the semisimple case applied to
$so(3)$, respectively $so(2,1)$ gives
\bea
E_2= 1 \oplus d_3 F_3\oplus F_3,
\eea
with
\begin{align}
F_3 &= \{ \hat\eta^3\,Q_1(f_1)\},\\
d_3 F_3&= \{ f_1\,\tilde Q_0(f_1)\}.
\end{align}

The associated elements of $H(\gamma\vert d, {\cal
B})$ are listed in table 2,  which involves the following new shorthand
notations
\begin{align}
\hat\eta^2 &= -\frac{1}{2}\,\epsilon_{abc}\,\eta^{a}\eta^b B^c,
\\ \hat\eta^1 &= \eta^{a}\,(G_a-\frac{1}{2}\epsilon_{abc} B^b B^c),
\\ \hat\eta^0 &= B^b G_b -\frac{1}{3!} \epsilon_{abc}B^{a}B^b B^c=
g_{ab}B^{a}dB^b + \frac{1}{3} \epsilon_{abc}B^{a}B^b B^c. \label{6.40}
\end{align}

\begin{table}[tbp]

\begin{tabular}{|c|c|c|c|c|}
\hline
 gh &  &$H(\gamma\vert d, {\cal B})$ &  & \\
\hline\hline
$0$ & 1 &$GA\, R_0(f_1,f_3)$ &0 & $\hat\eta^0 Q_1(f_1)$ \\
\hline
$1$ & $GC R_0(f_1,f_3)$& 0 & $\hat\eta^1 Q_1(f_1)$ & 0 \\ \hline
$2$ & 0 & $\hat\eta^2 Q_1(f_1) + \frac{1}{2} AC^2 T_1(f_2,f_3)$ & 0 & 0 \\
\hline
$3$ & $\hat\eta^3 Q_1(f_1) + \hat C^3 T_0(f_2,f_3)$ & $(\hat\eta^2 GC +
\hat\eta^3 GA)\, R_1(f_1,f_3)$ & 0 & 0\\
\hline
$4$ & $\hat\eta^3 GC R_1(f_1,f_3)$ & 0 & 0 & 0\\ \hline
$5$ & 0 & $(\hat\eta^2 \hat C^3 +\frac{1}{2}\hat\eta^3 AC^2)\,T_1(f_2,f_3)$ &
0 & 0\\
\hline
$6$ & $\hat\eta^3 \hat C^3 T_1(f_2,f_3)$ & 0 & 0 & 0 \\ \hline
\end{tabular}\\[1ex]
\centerline{\emph { Table 2}}
\end{table}

\subsection{$H(\gamma|d,\cA)$}

Using \eqref{gmoddA}, respectively \eqref{gmoddAn},
we have, for $0\leq p<n$,
\bea
H^p(\gamma|d,\cA)&\simeq& i_0 H^p(\gamma,\cB_\cK)\oplus F_1
\oplus \cN^p_\cA \oplus \cD^{-1}
(\cD H)^{p-1} (\gamma|d,\cB_\cK)\oplus\lambda^{\#}F^p_1,
\eea
and in form degree $n$,
\bea
H^n(\gamma|d,\cA)&\simeq & \cF^n\oplus  \cD^{-1} (\cD
H)^{n-1} (\gamma|d,\cB_\cK)\oplus\lambda^{\#}F^n_1,
\eea
with $\cF^n \simeq H^n(\gamma,\cA)/ d_0\cN^{n-1}_\cA$.

\section{Application to the consistent deformations of
2+1 dimensional gravity}

\subsection{Generalities}

$2+1$ dimensional gravity with vanishing cosmological constant
$\lambda$ is equivalent to a Chern-Simons theory based on the gauge
group $ISO(2,1)$ \cite{Achucarro:1986vz,Witten:1988hc}.

The Lie algebra
$iso(2,1)$ is not reductive and its Killing metric $G_{AB} =
f^D_{\,\,\,AC}\,f^C_{\,\,\,BD}$ is degenerate
\begin{equation} G_{AB} = \begin{pmatrix} g_{ab} &0\\ 0&
0\end{pmatrix},
\end{equation}
where
$g_{ab}$ is the Killing metric of the semi-simple $so(2,1)$ subalgebra.
However in this case, another invariant, symmetric and non degenerate metric
$\Omega^{(0)}_{AB}$ exists which allows for the construction of the CS
Lagrangian. The invariant quadratic form of interest is
\begin{equation}
 \Omega^{(0)}_{AB} = \begin{pmatrix}<J_a,J_b>& <J_a,P_b>\\
<P_a, J_b> & <P_a, P_b>\end{pmatrix} = \begin{pmatrix} 0 &
g_{ab}\\ g_{ab} & 0\end{pmatrix}.\end{equation}

Locally, the relation between
$2+1$ dimensional gravity and the Chern-Simons theory
is based on the $iso(2,1)$ Lie algebra valued one form
\begin{equation} {\cal A}_\mu = A_\mu^{A}\,T_A = e_\mu^{a}\,P_a +
\omega_\mu^{a}\,J_a\end{equation}
built from the dreibein fields
$e_\mu^{a}$ and the spin connection $\omega_\mu^{a}=\half
\epsilon^{a}_{bc}\omega_\mu^{bc}$ of $3$ dimensional Minkowski
spacetime $\cM$ with metric that we choose of signature $(-,+,+)$.
In terms of these variables,
the Chern-Simons action takes the form of the $2+1$ dimensional
Einstein-Hilbert action in vielbein formulation:
\begin{align} S^{(0)}_{CS} &= \int_{\cal M}\,
\Omega^{(0)}_{AB}\,[\frac{1}{2}\,A^{A}\,dA^B +
\frac{1}{6}\,A^A\,f^{B}_{\,\,\,CD}\,A^{C}A^D]
\label{CSact}\\ &=
\int_{\cal M}\,\frac{1}{2}\,(e^{a}\,d\,\omega_{a} +\omega^{a}\,de_a
+
\epsilon_{abc}\,e^{a}\,\omega^b\,\omega^c\,),\nonumber\\
&= \int_{\cal M}\,e^{a}\,G_a +
\frac{1}{2}\,d(e^{a}\,\omega_a).\label{lambda0}\end{align}

The gauge transformations are parametrized by two zero-forms
$\epsilon^{a}$ and $\tau^{a}$,
\begin{equation}\varepsilon =\varepsilon^AT_A=
\epsilon^{a}\,P_a + \tau^{a}\,J_a.
\end{equation}
Explicitly,
\begin{align}\delta_{\epsilon}\,e^{a} &= -\,d\epsilon^{a} -
\epsilon^{a}_{\,\,\,bc}\,(\omega^b\,\epsilon^c + e^b\,\tau^c),\\
\delta_{\epsilon}\,\omega^{a} &= -\,d\tau^{a} -
\epsilon^{a}_{\,\,\,bc}\,\omega^b\,\tau^c,
\end{align} and are equivalent, on shell, to local diffeomorphisms and
local Lorentz rotations. The classical equations of motion
express the vanishing of the field strenghts two-forms
\begin{align} F_a &= \frac{1}{2} F_{\mu\nu a}dx^\mu dx^\nu = de_a +
\epsilon_{abc}\,\omega^b\,e^c \\  G_a &= \frac{1}{2} G_{\mu\nu
a} dx^\mu dx^\nu = d\omega_a +
\frac{1}{2}\,\epsilon_{abc}\,\omega^b\,\omega^c
\end{align} where  \begin{equation} {\cal F}_{\mu\nu} = \cF_{\mu\nu}^{A}\,T_A
= F_{\mu\nu}^{a}\,P_a + G_{\mu\nu}^{a}\,J_a.\end{equation}
For invertible dreibeins, the equation $F^{a}=0$ can be algebraically
solved for $\omega^{a}$ as a function of the
$e^{a}$'s~; when substituted into the
remaining equation it tells that the space-time Riemann curvature
vanishes, which in $3$ dimensions implies that space-time is locally flat.

Our aim is to study systematically all consistent deformations of
$2+1$ dimensional gravity. By consistent, we mean
deformations of the action by local functionals and simultaneous
deformations of the gauge transformations such that the deformed
action is invariant under the deformed gauge transformations.

The problem of such consistent deformations can be reformulated
\cite{Barnich:1993vg} (for a review see \cite{Henneaux:1997bm}) as the
problem of deformations of the solution of the master equation and is
controled to first order by the cohomology $H^{0,3}(s|d)$.

\subsection{Results on local BRST cohomology}

The analysis of $H(s|d)$ for the Chern-Simons case (see e.g.
\cite{Barnich:2000zw}, section 14) implies that the cohomology
$H(s|d)$ is essentially given by the bottoms $[\half\hat\eta^3]$,
$[\hat C^3]$, $[\half\hat\eta^3\hat C^3]$ of $H(\gamma,\cB)$ and by their
lifts, which are all non trivial and unobstructed. (The only
additional classes correspond to the above bottoms multiplied
by non exact spacetime forms.)

It follows that
$H^{0,3}(s|d)$ is obtained
from the lift (associated to $s$) of the elements $[\half\hat\eta^3]$ and
$[\hat C^3]$ of $H^{3,0}(s)$.  The former element can be lifted to
$\half \hat\eta^0$ with $\hat\eta^0$ given in \eqref{6.40}.  It
corresponds to the Chern-Simons action built on $so(2,1)$.
The results on $H(\gamma|d,\cB)$ (see table 2) imply that
the lift of $\hat C^3$ in $H(s|d)$ cannot been done without a non
trivial dependence on the antifields, and hence without a non trivial
deformation of the gauge transformations. Following again
\cite{Barnich:2000zw}, this lift is given by
\bea
\epsilon_{abc}\,[\frac{1}{6}\,e^{a}\,e^b\,e^c + e^{a}\,\star
\omega^{*b}\,C^c + \frac{1}{2}\,\star \eta^{*a}\,C^b\,C^c\,],
\eea
where $\star\omega^{*a} =
\frac{1}{2}dx^\mu dx^\nu
\epsilon_{\mu\nu\rho}\,\omega^{*a\rho}$ and $\star \eta^*_a =
d^3x\,\eta^*_a$.
The antifield independent part gives the deformation of the original
action.

Note also that $H^{1,3}(s|d)$ is trivial, which implies that there can
  be no anomalies in a perturbative quantization of $2+1$ dimensional
  gravity. Furthermore, the starting point Lagrangian $3$ form $eG$ is
  trivial, $[eG]=[0]\in H^{0,3}(s|d)$, which is the reason why we do
  not introduce a separate coupling constant for this term.

\subsection{Maximally deformed $2+1$ dimensional gravity}

Introducing coupling constants $\mu$ and $\lambda$ (the cosmological constant)
for the $2$ first order deformations, they can be easily shown to
extend to all orders by introducing a $\lambda\mu$ dependent term in
the action. The associated completely deformed theory can be written
as a Chern-Simons theory in terms of the deformed invariant metric
\bea
\Omega^{\lambda,\mu}_{AB}=\Omega^{(0)}_{AB}+G^{\mu,\lambda}_{AB},\\
G^{\mu,\lambda}_{AB}=
\mu \begin{pmatrix}g_{ab}&0\\0&\lambda\,g_{ab}\end{pmatrix},
\eea
and the deformed structure constants $f^{A(\lambda)}_{\,\,\,BC}$
given by
\begin{equation} [J_a,J_b] = \epsilon_{abc}\,J^c,\,[J_a,P_b] =
\epsilon_{abc}\,P^c,\,[P_a,P_b] =
\lambda\,\epsilon_{abc}\,J^c.\end{equation}
For $\lambda>0$, these structure constants are those of the
semi-simple Lie algebra $so(2,1)\oplus so(2,1)$.
The deformed Chern-Simons action reads explicitly
\begin{align}
S^{\lambda,\mu}&=\int_{\cal M}\,
\Omega^{\lambda,\mu}_{AB}\,[\frac{1}{2}\,A^{A}\,dA^B +
\frac{1}{6}\,A^A\,f^{B(\lambda)}_{\,\,\,CD}\,A^{C}A^D]\\ &=
\int_{\cal
M} \Big[e^a G_a +\mu \,[\frac{1}{2}\,\omega^{a}\,d\omega_a +
\frac{1}{6}\,\epsilon_{abc}\,\omega^{a}\,\omega^{b}\,\omega^c]\nonumber\\
&+\lambda\,\frac{1}{3!}\,\epsilon_{abc}\,e^{a}\,e^b\,e^c\,
+
\lambda\mu\,\,[\frac{1}{2}\,e^{a}\,de_a +
\frac{1}{2}\,\epsilon_{abc}\,e^{a}\,e^{b}\,\omega^c]\Big],
\label{actlambdamu}
\end{align}
while the deformed gauge transformations read
\begin{align}\delta_\epsilon\,e^{a} &= -\,d\epsilon^{a} -
\epsilon^{a}_{\,\,\,bc}\,(\omega^b\,\epsilon^c + e^b\,\tau^c),\\
\delta_\epsilon\,\omega^{a} &= -\,d\tau^{a} -
\epsilon^{a}_{\,\,\,bc}\,(\omega^b\,\tau^c +\lambda\,e^b\,\epsilon^c).
\end{align}

Thus, our analysis shows that there are no other consistent
deformations of $2+1$ dimensional gravity than those already discussed
in \cite{Witten:1988hc}.

\section*{Acknowledgments}

The authors want to thank M.~Henneaux for suggesting the problem and
for useful discussions. Their work is supported in part by the
``Actions de Recherche Concert\'ees'' of the ``Direction de la Recherche
Scientifique-Communaut\'e Fran\c caise de Belgique, by a ``P\^ole
d'Attraction Interuniversitaire'' (Belgium), by IISN-Belgium
(convention 4.4505.86), by Proyectos FONDECYT 1970151 and 7960001
(Chile) and by the European Commission RTN programme HPRN-CT00131, in
which they are associated to K.~U.~Leuven.

\section*{Appendix A: Descents and decomposition according to
  homogeneity}
\setcounter{equation}{0}
\def\theequation{A.\arabic{equation}}

The space $\cE$ can be decomposed into monomials of definite homogeneity $M$ in
the fields and their derivatives, $\cE=\oplus_{M=0}\cE_M$, and one can define
the spaces of polynomials of homogeneity greater or equal to $M$, $\cE^M=
\oplus_{N\geq M} \cE_N$.

If
\bea
\cC^M=<H(\gamma|d,\cE^M ),H(\gamma,\cE^M ),{\cal D}^M,{l^{\#}}^M,{i^{\#}}^M>,\\
\cC_M=<H(\gamma_0|d,\cE_M ),H(\gamma_0,\cE_M ),{\cal D}_M,l^{\#}_M,i^{\#}_M>,
\eea
are the exact couples that describe the descents of $\gamma$ in $\cE^M$,
respectively of $\gamma_0$ in $\cE_M$, one can define mappings
between exact couples (see e.g.
\cite{Hubook}) through
\bea
I_{M+1}=(j_{M+1},k_{M+1}): \cC^{M+1}\longrightarrow  \cC^M,\\
P_M=(\pi_M,\psi_M):  \cC^M\longrightarrow \cC_M,\\
G_M=(m_M,n_M):  \cC_M  \longrightarrow \cC^{M+1}.
\eea
The map $j_{M+1}$ consists in the natural injection of elements of
$H(\gamma|d,\cE^{M+1})$ in $H(\gamma|d,\cE^{M})$ and similarily $k_{M+1}$
consists in the injection of elements of $H(\gamma,\cE^{M+1})$ as elements of
$H(\gamma,\cE^{M})$, with
\bea
\cD^M\circ j_{M+1}=j_{M+1}\circ \cD^{M+1},\\
{l^{\#}}^M\circ j_{M+1} =k_{M+1} \circ {l^{\#}}^{M+1},\\
 {i^{\#}}^M\circ k_{M+1} =j_{M+1} \circ{i^{\#}}^{M+1}.
\eea
The map $\pi_M: H(\gamma|d,\cE^M )\longrightarrow H(\gamma_0|d,\cE_M )$ is
defined by $\pi_M[A^M]= [A_M]$, while $\psi_M: H(\gamma,\cE^M )\longrightarrow
H(\gamma_0,\cE_M )$ is defined by $\psi_M[a^M]= [a_M]$. Again, the various maps
commute,
\bea
 \cD_M\circ \pi_{M}=\pi_{M}\circ \cD^{M},\\
{l^{\#}}_M\circ \pi_{M} =\psi_{M} \circ {l^{\#}}^{M},\\
i^{\#}_M\circ \psi_{M} =\pi_{M} \circ {i^{\#}}^{M}.
\eea
Both the maps $m_M: H(\gamma_0|d,\cE_M )\longrightarrow
H(\gamma|d,\cE^{M+1})$ and $n_M:H(\gamma_0,\cE_M)\longrightarrow
H(\gamma,\cE^{M+1}$ are defined by are defined by the induced
action of $\gamma_1$: $m_M [a_M]=[\gamma_1 a_M]$ and $n_M[a_M
]=[\gamma_1 a_M]$, with
\bea
\cD^{M+1}\circ m_{M}=m_{M}\circ \cD_{M},\\
{l^{\#}}^{M+1}\circ m_{M} =n_{M} \circ l^{\#}_{M},\\
{i^{\#}}^{M+1}\circ n_{M} =m_{M} \circ i^{\#}_{M}.
\eea
Finally, the triangles
\begin{eqnarray}
\begin{array}{ccc} H(\gamma|d,\cE^{M+1} )\stackrel{j_{M+1}}{\longrightarrow}
H(\gamma|d,\cE^{M} )\\ m_M\nwarrow\ \swarrow \pi_M\\ H(\gamma_0|d,\cE_M )
\end{array}\label{et1},
 \end{eqnarray}
\begin{eqnarray}
\begin{array}{ccc} H(\gamma,\cE^{M+1} )\stackrel{k_{M+1}}{\longrightarrow}
H(\gamma,\cE^{M} )\\ n_M \nwarrow\ \swarrow \psi_M \\ H(\gamma_0,\cE_M)
\end{array}\label{et2},
\end{eqnarray}
are exact at all corners, implying, if $j_0=1=k_0$, that
\bea
H(\gamma|d,\cE)=\oplus_{M=0}^\infty j_0\dots j_M\pi_M^{-1}{\rm Ker}\ m_M,\\
H(\gamma,\cE)=\oplus_{M=0}^\infty k_0\dots k_M\psi_M^{-1}{\rm Ker}\ n_M.
\eea

All this can be summarized by the commutative diagram of
figure 1. The corners of the big triangle are itself given by exact
triangles and the large triangles obtained by taking a group at
the same position of each small triangle are also exact.

\begin{figure}[h]
\caption{Exact triangle of exact triangles}
\centerline{\includegraphics[scale=0.9]{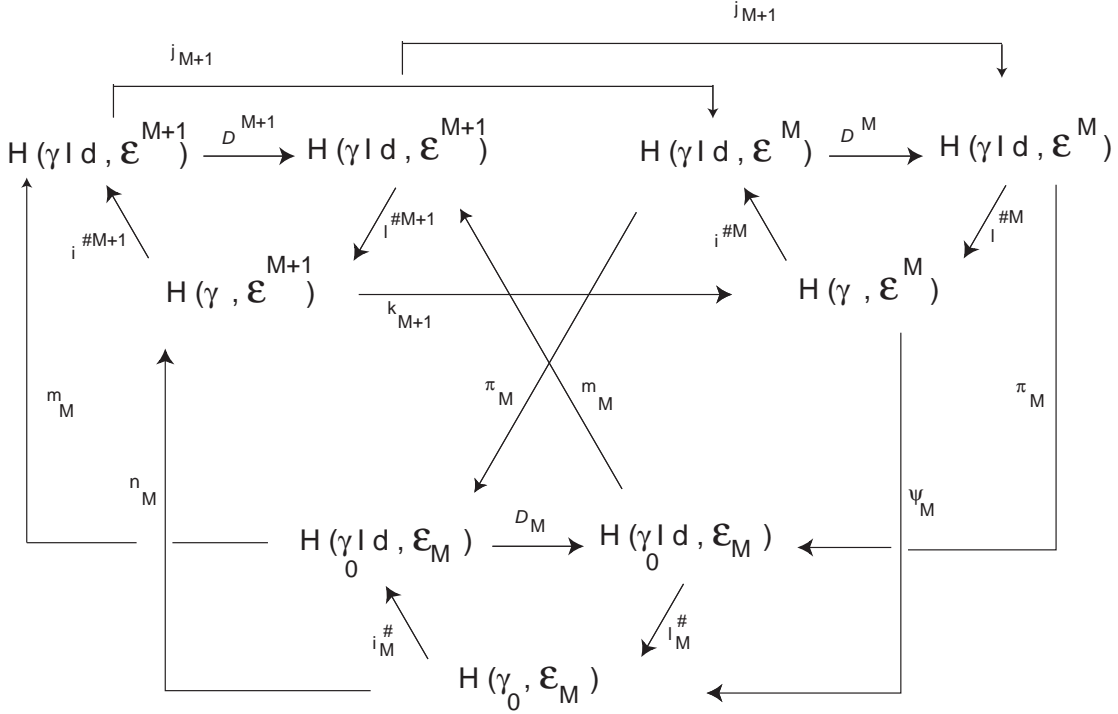}}
\end{figure}

\newpage


\providecommand{\href}[2]{#2}\begingroup\raggedright\endgroup

\end{document}